\documentclass[journal]{IEEEtran}
\usepackage[utf8]{inputenc}
\usepackage{amsmath,amssymb,mathtools}
\usepackage{graphicx}
\usepackage{booktabs,multirow,threeparttable,adjustbox}
\usepackage{cite}
\usepackage{url}
\usepackage{balance}
\usepackage{hyperref}
\usepackage{algorithm}
\usepackage{algpseudocode}
\usepackage{tikz}
\usetikzlibrary{shapes.geometric, arrows.meta, positioning, arrows}

\begin{document}

\title{An Accurate Modulation-Assisted Method for Phasor Estimation Under Decaying DC Offsets
\vspace{-2mm}}

\author{Sina~Mohammadi,~\IEEEmembership{Graduate Student Member,~IEEE,}
        Apurv~Shukla,~\IEEEmembership{Member,~IEEE,}
        \\Junho~Hong,~\IEEEmembership{Senior Member,~IEEE,}
        Wencong~Su,~\IEEEmembership{Senior Member,~IEEE}
\vspace{-12mm}        
\thanks{S. Mohammadi, A. Shukla, J. Hong and W. Su are with the Department
of Electrical and Computer Engineering, University of Michigan-Dearborn, Dearborn, MI, 48128, USA. (e-mail: sinamo@umich.edu; apurvshu@umich.edu; jhwr@umich.edu; wencong@umich.edu) (corresponding author: Wencong Su)}
\vspace{-0.1mm}
}

\maketitle

\begin{abstract}

Decaying DC (DDC) offsets severely compromise the accurate estimation of current phasors immediately following power grid fault occurrence. Consequently, conventional full-cycle discrete Fourier transform (FCDFT) algorithms experience decaying oscillatory magnitude errors, potentially causing distance relays to malfunction. This paper introduces a modulation-assisted technique designed to precisely eliminate the DDC components from fault current signals. By multiplying the fault signal with a high-frequency carrier and applying consecutive sliding Fourier projections, the proposed method effectively isolates the DDC signal while entirely rejecting higher-order harmonic interference. It utilizes a direct algebraic solution derived from a robust forward-difference operation across just three sequential projections to estimate and extract the DDC parameters. Furthermore, the algorithm seamlessly processes both single and multiple DDC components by dynamically tracking them as one equivalent first-order mode. Extensive evaluations are conducted to validate the proposed method accuracy. The proposed algorithm limits the maximum overshoot (MO) to 0.62\% (0.59\% for $k >= 8$) for the evaluated DDC cases, and to 2.47\% (0.95\% at 20 dB for $k = 8$) under noisy conditions.

\end{abstract}

\vspace{-3mm}

\vspace{-5mm}
\section{Introduction}
\label{sec:introduction}

\IEEEPARstart{F}{ast} and accurate estimation of the fundamental fault-current phasor is a basic requirement of modern power-system protection. Distance, directional, differential, and overcurrent relays use current phasors to identify the faulted element and issue a secure trip command within a few cycles. The discrete Fourier transform (DFT) is widely employed for this purpose because it has a simple implementation, a fixed computational burden, and exact rejection of integer harmonics under nominal-frequency steady-state conditions. However, its performance deteriorates during the first few cycles following a transmission line fault. Fault currents generally contain harmonics, measurement noise, and one or more exponentially decaying dc (DDC) components. Because a DDC is nonperiodic within the DFT window, it is not orthogonal to the fundamental-frequency basis functions and produces a decaying oscillatory error in the estimated magnitude and phase \cite{guo2003,kang2009,mohammadi2022review,mohammadi2022novel,mohammadi2026symbolicphasor}. This transient error can delay relay convergence and cause relay overreach or underreach.

The DDC observed by a protection relay may have more than one physical origin. The primary component is produced by the transient response of the faulted power network; its initial magnitude depends mainly on the fault-inception angle, whereas its time constant is governed by the post-fault resistance-to-inductance ratio. A second component may be introduced by the transient response of the current transformer (CT). The primary and CT-induced components can have substantially different amplitudes and decay rates, particularly during high-current faults and CT saturation. The problem is also time critical: the DDC may persist beyond the decision interval of a high-speed relay, while waiting for its natural decay defeats the purpose of fast phasor estimation \cite{mohammadi2022review,afrandideh2022,jamali2022}. An effective estimator should therefore suppress both single- and multiple-DDC components without sacrificing harmonic rejection, noise tolerance, or convergence speed.

Existing DDC-mitigation methods can be broadly classified according to whether they preprocess the current waveform, correct the DFT output, or explicitly identify multiple exponential modes. In the first class, the DDC is suppressed before phasor calculation. The digital mimic filter is a representative solution and can substantially reduce the offset before the current is applied to the DFT \cite{benmouyal1995}. Nevertheless, its response depends on the agreement between the selected and actual DDC time constants, which vary with the fault condition. In addition, it may also amplify noise and high-frequency transients. Adaptive mimic filtering reduces the time-constant-mismatch problem but introduces an additional parameter-adaptation stage \cite{yu2006discrete}. Other preprocessing approaches use empirical-mode decomposition, intrinsic time-scale decomposition, recursive wavelets, or complex-frequency filters \cite{jiang2014modified,pazoki2017new,ren2009,jafarpisheh2018phasor}. Although these methods can attenuate DDC effect before phasor extraction, their decomposition, reconstruction, or auxiliary-filtering requirements are less direct than those of a conventional DFT relay.

The second class modifies the DFT calculation or estimates the leakage introduced by an assumed DDC model. Simplified Fourier-domain corrections use relations among consecutive samples or DFT outputs to compensate for the exponential term \cite{guo2003,kang2009}. Related modified-DFT formulations combine half- or full-cycle outputs, odd and even sample sets, or selected spectral components to suppress the offset-induced error \cite{silva2017modified}. Online parameter-estimation methods infer the decay factor and remove its contribution from the measured phasor \cite{altallaq2011}, while analytical and fast-removal formulations seek to shorten the transient interval using compact sample combinations \cite{mohammadi2022novel,abdoos2016,hajizadeh2020analytical}. High-frequency modulation has also been used to move DDC information away from zero frequency and then recover the fundamental phasor through modified DFT operations \cite{afrandideh2022,7745975}. These methods can significantly improve the conventional DFT when the adopted signal model is satisfied. The method in \cite{liu2026analytical} segments a one-cycle window into subsets and applies DFT to construct a complex equation system incorporating the fundamental frequency phasor alongside two decaying DC offsets. However, the computational complexity is high for the algorithm.

Window length is a central design variable in these formulations. Subcycle estimators respond early but weaken noise averaging and may forfeit full-cycle harmonic rejection \cite{kumar2021novel}. Weighted least square and adaptive-window estimators provide flexibility at the cost of a larger signal model or parameter-selection mechanism \cite{jafarian2011weighted,hwang2018least,mahari2017adaptive,85841}. Conversely, full-cycle integration preserves useful Fourier properties but introduces observation delay. Taylor approximation simplifies the estimation equations, although its truncation error depends on the decay-time range and sampling interval. Frequency-deviation-aware formulations address another error source but require additional estimation stages \cite{tajdinian2016analytical}. These tradeoffs become more consequential when the primary and CT-induced offsets have widely separated decay rates \cite{afrandideh2022,jamali2022,achlerkar2019assessment}.

The explicit treatment of two or more DDC modes has consequently received increasing attention. A modulation-based DFT method estimates the primary and secondary DDC components by combining successive outputs of one-cycle averaging filters and removes their contributions from the conventional DFT result \cite{afrandideh2022}. A differential-equation formulation decomposes the fault current into modal components and permits the moving-window length to be selected according to the required compromise between accuracy and delay \cite{hajimohammadi2024}. Selective interval integration estimates the system and CT offsets from two properly separated integration intervals without applying a Taylor approximation to the exponential terms \cite{jamali2022}. Multiple exponentially decaying components have also been addressed directly in the DFT domain \cite{parpaei2022,hwang2018dft}. Half-cycle and hybrid formulations estimate two exponential modes through auxiliary signals, successive integrations, or carrier modulation \cite{tajdinian2017half,allahbakhshi2019hybrid}. These approaches establish that retaining separate decay modes can eliminate the residual oscillations that remain after an equivalent single-DDC approximation. At the same time, their implementations involve different combinations of auxiliary filtering, interval selection, successive-window processing, model assumptions, or numerical parameter extraction. Thus, there remains a need for a concise formulation that preserves the harmonic selectivity of a full-cycle DFT, without requiring prior knowledge of the number of DDC components or iterative nonlinear parameter estimation.

This paper develops a  modulation-assisted estimator to meet protective relaying requirements. The measured current is multiplied by a high-frequency carrier signal, which shifts the non-oscillatory DDC component into damped components centered at the selected carrier frequency. A sliding full-cycle projection is then applied to reject standard integer harmonics. The high-frequency modulation removes the contribution of the oscillatory term, allowing the projection sequence to be approximated by a decaying exponential. By shifting these consecutive projections by a small fractional time step, the outputs form a predictable discrete sequence. Unlike conventional estimators that rely on iterative approximations, the proposed sequence is resolved purely algebraically. Applying a simple forward difference operation across consecutive observations entirely removes constant residual integration errors. The resulting difference equation directly yields the precise decay characteristics, allowing the algorithm to dynamically track the DDC component and deliver a fully compensated fundamental phasor estimate continuously. The contributions of this paper are as follows:

\begin{enumerate}
\item A robust high-frequency modulation technique is introduced to decouple DDC offsets from fundamental fault phasors. By translating non-oscillatory components into a higher spectrum, a full-cycle sliding projection precisely isolating pure DDC terms without iterations.

\item A flexible modulation multiplier selection framework optimizes noise immunity and minimizes estimation transients. This adaptive approach guarantees overall hardware compatibility, seamlessly accommodating both standard relay anti-aliasing filter bandwidth limits and diverse digital distance protection sampling rates.

\item A direct algebraic forward-difference formulation dynamically extracts equivalent DDC parameters. This robust first-order model accurately approximates the composite transient behaviors, seamlessly scaling to resolve multiple interfering DDC components.
\end{enumerate}

The remainder of this paper is organized as follows. Section II presents the proposed algorithm for eliminating the DDC component via high-frequency modulation, detailing both the time- and frequency-domain derivations. Section III evaluates the performance of the proposed method under various scenarios, including single and double DDC components, measurement noise, varying sampling rates, odd harmonics, and off-nominal frequencies, followed by simulation-based assessments on a test network. Finally, Section IV concludes the paper and summarizes the key capabilities of the method.

\vspace{-3mm}
\section{Proposed Algorithm}

\subsection{Modulation-Assisted DDC Parameter Extraction}

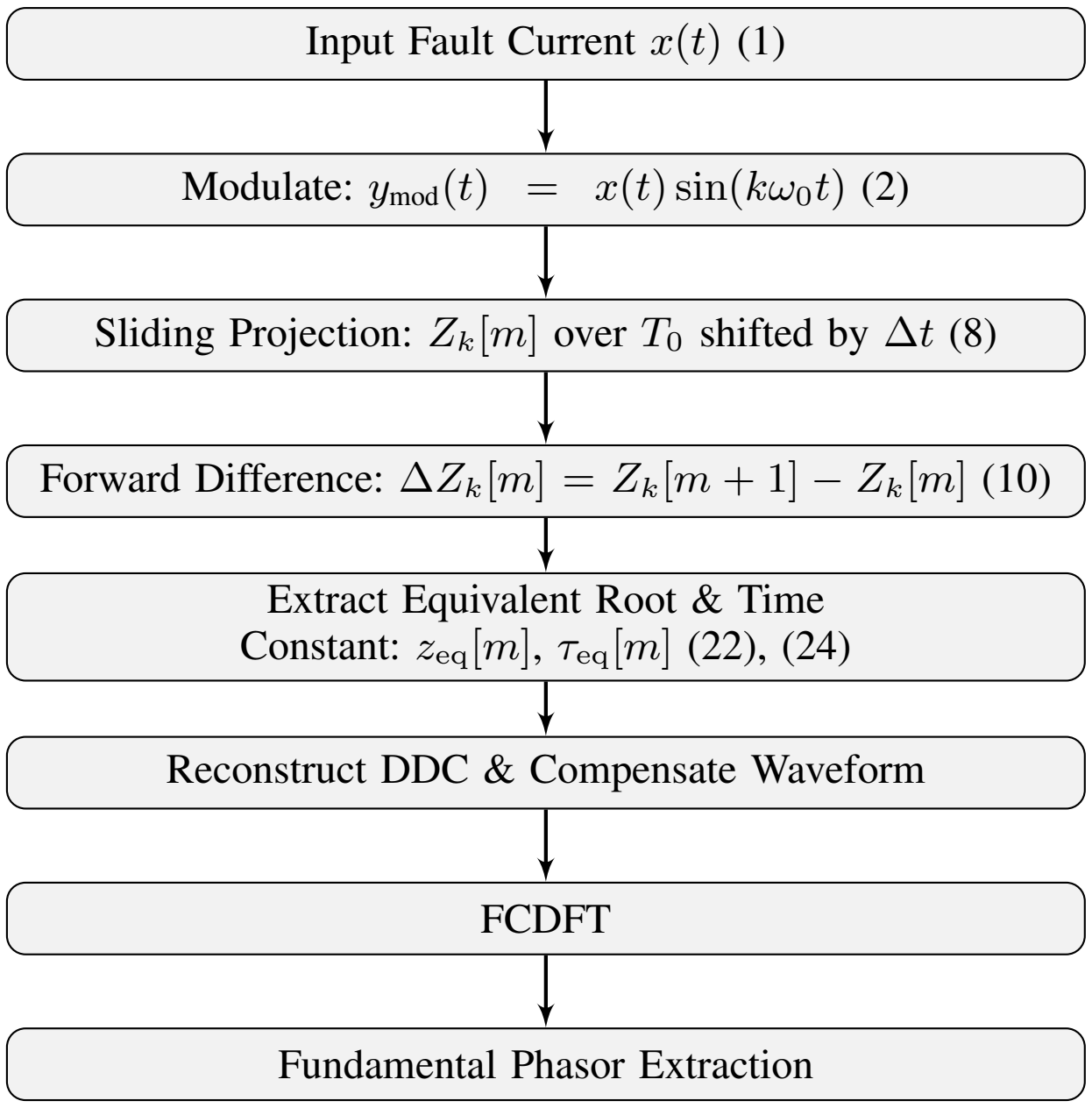
\begin{figure}[htbp]
\centering
\scriptsize
\tikzstyle{block} = [rectangle, draw, fill=gray!10, text width=0.9\columnwidth, text centered, rounded corners, minimum height=2.2em, font=\small]
\tikzstyle{line} = [draw, -latex', thick]
\begin{tikzpicture}[node distance=1.1cm, auto]
    \node [block] (input) {Input Fault Current $x(t)$ \eqref{eq:single_ddo_signal}};
    \node [block, below of=input] (mod) {Modulate: $y_{\text{mod}}(t) = x(t) \sin(k\omega_0 t)$ \eqref{eq:modulated_signal}};
    \node [block, below of=mod] (proj) {Sliding Projection: $Z_k[m]$ over $T_0$ shifted by $\Delta t$ \eqref{eq:Zk_simplified_final}};
    \node [block, below of=proj] (diff) {Forward Difference: $\Delta Z_k[m] = Z_k[m+1] - Z_k[m]$ \eqref{eq:homo_1ddo}};
    \node [block, below of=diff] (root) {Extract Equivalent Root \& Time Constant: $z_{\mathrm{eq}}[m]$, $\tau_{\mathrm{eq}}[m]$ \eqref{eq:zeq_ratio_derivation}, \eqref{eq:tau_eq_dynamic}};
    \node [block, below of=root] (clean) {Reconstruct DDC \& Compensate Waveform};
    \node [block, below of=clean] (dft) {FCDFT};
    \node [block, below of=dft] (final) {Fundamental Phasor Extraction};
    
    \path [line] (input) -- (mod);
    \path [line] (mod) -- (proj);
    \path [line] (proj) -- (diff);
    \path [line] (diff) -- (root);
    \path [line] (root) -- (clean);
    \path [line] (clean) -- (dft);
    \path [line] (dft) -- (final);
\end{tikzpicture}
\caption{Flowchart of the proposed modulation-assisted method for removing DDC of the fault current.}
\label{fig:flowchart}
\vspace{-5mm}
\end{figure}

The overall schematic diagram of the proposed method is summarized in the flowchart presented in Fig.~\ref{fig:flowchart}. The proposed method consists of signal modulation, consecutive sliding projections, a forward-difference operation, DDC reconstruction, and finally the fundamental phasor estimation.

\subsubsection{Signal Modulation and Sliding Projection Derivation}
Consider a fault current signal containing a single DDC component alongside $N$ harmonics:
\begin{equation} \label{eq:single_ddo_signal}
    x(t) = I_0 e^{-t/\tau} + \sum_{n=1}^N I_n \cos(n\omega_0 t + \theta_n)
\end{equation}
where $I_0$ and $\tau$ are the initial amplitude and time constant of the DDC, respectively, $\omega_0 = 2\pi/T_0$ is the fundamental angular frequency, and $I_n$ and $\theta_n$ represent the amplitude and phase angle of the $n$-th harmonic component.

Modulating $x(t)$ with a harmonic carrier $\sin(k\omega_0 t)$, where $k$ is the integer harmonic carrier multiplier, yields:
\begin{equation} \label{eq:modulated_signal}
    y_{\text{mod}}(t) = I_0 e^{-t/\tau}\sin(k\omega_0 t) + \sum_{n=1}^N I_n \cos(n\omega_0 t + \theta_n)\sin(k\omega_0 t)
\end{equation}
A sliding-window Fourier projection is applied specifically at the $k$-th harmonic bin over a fundamental cycle $T_0$, starting at an arbitrary initial time $t_0 = m\Delta t$ for the $m$-th observation window, where $\Delta t$ is the sliding time step. Assuming the fault current does not contain a significant harmonic of order $n=2k$, the cross-product terms of integer harmonics integrate precisely to zero over $T_0$, simplifying the complex projection to:
\begin{equation} \label{eq:sliding_integral_base}
    Z_k[m] = \int_{t_0}^{t_0 + T_0} I_0 e^{-t/\tau} \sin(k\omega_0 t) e^{-j k \omega_0 t} \, dt
\end{equation}
Applying Euler's identity, the mixed trigonometric-exponential term expands into a stationary baseband term and a double-frequency oscillatory term:
\begin{equation} \label{eq:euler_expansion}
    \sin(k\omega_0 t) e^{-j k \omega_0 t} = \frac{1 - e^{-j 2k \omega_0 t}}{2j}
\end{equation}
Substituting \eqref{eq:euler_expansion} into \eqref{eq:sliding_integral_base} splits the projection into two distinct integrals:
\begin{equation} \label{eq:Zk_split_integrals}
    Z_k[m] = \frac{I_0}{2j} \underbrace{\int_{t_0}^{t_0 + T_0} e^{-t/\tau} \, dt}_{\text{Pure Decay Term}} - \frac{I_0}{2j} \underbrace{\int_{t_0}^{t_0 + T_0} e^{-t\left(\frac{1}{\tau} + j 2k\omega_0\right)} \, dt}_{\text{Oscillatory Term}}
\end{equation}
Simplifying both integrals at the boundaries $t_0$ and $t_0 + T_0$ yields:
\begin{equation} \label{eq:Zk_evaluated_limits}
    Z_k[m] = \frac{I_0}{2j} \left(1 - e^{-T_0/\tau}\right) e^{-t_0/\tau} \left[ \tau - \frac{e^{-j 2k\omega_0 t_0}}{\frac{1}{\tau} + j 2k\omega_0} \right]
\end{equation}
where the simplification $e^{-j 2k\omega_0 T_0} = 1$ has been applied. 

The significance of the oscillatory phase term relative to the pure decay term $\tau$ is given by the ratio $\eta$:
\begin{equation} \label{eq:error_ratio}
    \eta = \frac{\left| \frac{e^{-j 2k\omega_0 t_0}}{1/\tau + j 2k\omega_0} \right|}{\tau} = \frac{1}{\sqrt{1 + (2k\omega_0 \tau)^2}}
\end{equation}
For practical power system faults ($0.5 T_0 \le \tau \le 5 T_0$ \cite{benmouyal1995}), selecting a sufficiently high carrier multiplier ($k \ge 4$) guarantees the condition $k\omega_0 \gg 1/\tau$, making the oscillatory term mathematically negligible ($\eta < 0.04$). However, the maximum value of $k$ is practically constrained by the digital relay's hardware configuration. To prevent signal degradation, the translated carrier frequency ($k f_0$, where $f_0 = 1/T_0$ is the fundamental frequency) must remain strictly below the cut-off frequency ($f_c$) of the relay's analog anti-aliasing filter (AAF) and strictly within the digital Nyquist limit \cite{5235740}. Therefore, an optimal modulation index (e.g., $4 \le k \le 8$ for standard commercial relays) is selected to simultaneously eliminate the oscillatory integration error while preserving the integrity of the isolated DDC component within the AAF's flat passband.

Omitting this negligible component, the complex sliding projection is reduced to a purely decaying sequence without phase oscillations:
\begin{equation} \label{eq:Zk_simplified_final}
    Z_k[m] \approx \left[ \frac{I_0 \tau}{2j} \left(1 - e^{-T_0/\tau}\right) \right] e^{-t_0/\tau} = \mathbf{K} e^{-m\Delta t/\tau}
\end{equation}
where $\mathbf{K}$ is a constant complex scalar, and the substitution $t_0 = m\Delta t$ discretizes the temporal decay.

\subsubsection{Single DDC}
When exactly one physical decaying mode is present, the discrete sequence modeled in \eqref{eq:Zk_simplified_final} follows a first-order linear homogeneous difference equation:
\begin{equation} \label{eq:base_1ddo}
    Z_k[m+1] + a_0 Z_k[m] = 0
\end{equation}
where $a_0 = -e^{-\Delta t/\tau_1}$ is the autoregressive coefficient. While $a_0$ could theoretically be extracted directly from the raw projections, practical fault currents contain low-frequency A/D offsets and unsuppressed spectral leakage. To remove gradual background noise and ensure computational stability, the difference between consecutive points is evaluated as $\Delta Z_k[m] = Z_k[m+1] - Z_k[m]$. The differenced sequence maintains the homogeneous property:
\begin{equation} \label{eq:homo_1ddo}
    \Delta Z_k[m+1] + a_0 \Delta Z_k[m] = 0
\end{equation}
Evaluating this at $m=0$ provides the robust coefficient $a_0$:
\begin{equation} \label{eq:a0_1ddo}
    a_0 = -\frac{\Delta Z_k[1]}{\Delta Z_k[0]} = -\frac{Z_k[2] - Z_k[1]}{Z_k[1] - Z_k[0]}
\end{equation}
The discrete root is $z_1 = -a_0$, and the time constant $\tau_1$ is extracted as:
\begin{equation} \label{eq:tau_1ddo}
    \tau_1 = \frac{-\Delta t}{\ln(z_1)}
\end{equation}
With $\tau_1$ known, the initial amplitude $I_{01}$ is recovered algebraically from the differenced projection components after compensating for the isolated exponential mode.

\subsubsection{Two DDC}
The fault current may exhibit two DDCs. In this case, the discrete sequence is modeled by a second-order linear homogeneous difference equation:
\begin{equation} \label{eq:base_2ddo}
    Z_k[m+2] + a_1 Z_k[m+1] + a_0 Z_k[m] = 0
\end{equation}
Similar to the single DDC scenario, applying the difference operator $\Delta Z_k[m] = Z_k[m+1] - Z_k[m]$ across consecutive equations filters out practical baseline offsets, transforming the system into a robust homogeneous form on the differences:
\begin{equation} \label{eq:homo_2ddo}
    \Delta Z_k[m+2] + a_1 \Delta Z_k[m+1] + a_0 \Delta Z_k[m] = 0
\end{equation}
To determine the unknown autoregressive coefficients ($a_0, a_1$), five consecutive sliding windows ($m \in \{0, 1, 2, 3, 4\}$) are processed to form a fully determined $2 \times 2$ linear system:
\begin{equation} \label{eq:matrix_2ddo_diff}
    \begin{bmatrix} \Delta Z_k[0] & \Delta Z_k[1] \\ \Delta Z_k[1] & \Delta Z_k[2] \end{bmatrix} \begin{bmatrix} a_0 \\ a_1 \end{bmatrix} = \begin{bmatrix} -\Delta Z_k[2] \\ -\Delta Z_k[3] \end{bmatrix}
\end{equation}
Solving this matrix yields $a_0$ and $a_1$. The discrete roots $z_1$ and $z_2$ are derived from the characteristic equation $z^2 + a_1 z + a_0 = 0$. The true time constants are subsequently calculated as $\tau_i = -\Delta t / \ln(z_i)$. The complex scalars $\mathbf{K}_1$ and $\mathbf{K}_2$, along with their corresponding initial amplitudes $I_{01}$ and $I_{02}$, are then isolated by fitting the discrete roots back to the differenced projection sequence $\Delta Z_k[m] = \sum_{i=1}^2 \mathbf{K}_i (z_i - 1) z_i^m$ to extract the exponential parameters while preserving baseline-offset rejection.

\subsubsection{Generalized Extraction ($n$-DDC)}
The formulation readily scales to resolve an arbitrary integer number $n$ of independent decaying modes. The projection sequence observes an $n$-th order linear recurrence relation:
\begin{equation} \label{eq:base_nddo}
    \sum_{p=0}^n a_p Z_k[m+p] = 0, \quad \text{where } a_n = 1
\end{equation}
Applying the difference operator $\Delta$ to essentially filter out low-frequency components yields a homogeneous relation on the differenced sequence:
\begin{equation} \label{eq:homo_nddo}
    \sum_{p=0}^n a_p \Delta Z_k[m+p] = 0, \quad \text{where } a_n = 1
\end{equation}
Parameterizing $n$ distinct modes demands an observation vector of $2n+1$ consecutive sliding windows (indices $m=0$ through $2n$). The autoregressive coefficients $\mathbf{a} = [a_0, \dots, a_{n-1}]^T$ are determined via the $n \times n$ Hankel matrix formulation:
\begin{equation} \label{eq:matrix_nddo_diff}
    \begin{aligned}
        &\begin{bmatrix}
        \Delta Z_k[0] & \cdots & \Delta Z_k[n-1] \\
        \vdots & \ddots & \vdots \\
        \Delta Z_k[n-1] & \cdots & \Delta Z_k[2n-2]
        \end{bmatrix}
        \begin{bmatrix} a_0 \\ \vdots \\ a_{n-1} \end{bmatrix} \\
        &\quad = -
        \begin{bmatrix} \Delta Z_k[n] \\ \vdots \\ \Delta Z_k[2n-1] \end{bmatrix}
    \end{aligned}
\end{equation}
The mode roots $\{z_1, \dots, z_n\}$ are identified as the roots of the $n$-th degree polynomial $\sum_{p=0}^n a_p z^p = 0$. Using these extracted roots, an $n \times n$ Vandermonde system \cite{bjorck1970solution} is constructed from the first $n$ differenced sequential projections $\Delta Z_k[m] = \sum_{i=1}^n \mathbf{K}_i (z_i - 1) z_i^m$ to jointly estimate the individual amplitude coefficients $\mathbf{K}_i$.


\vspace{-5mm}
\subsection{Equivalent Single-Mode Modeling for Multiple DDC}
While the previous subsections established the formulations for $n=1$, $n=2$, and a general $n$, the number of DDC components need not be known a priori because their combined effect is represented by a dynamically updated equivalent DDC component. Rather than implementing complex multi-order extraction routines, the combined effect of multiple decaying modes can be piecewise approximated by a single, dynamically updating equivalent decaying offset as demonstrated in \cite{kang2009} and \cite{silva2017modified}. To illustrate this concept, the remainder of this subsection demonstrates the equivalent modeling specifically for two DDCs.

Consider a discrete sliding-window sequence obtained from a fault current containing two distinct DDC components. Under the proposed framework, the complex projection expands into a linear combination of two geometric progressions:
\begin{equation} \label{eq:multi_base}
    Z_k[m] = \mathbf{K}_1 z_1^m + \mathbf{K}_2 z_2^m
\end{equation}
where $z_1 = e^{-\Delta t/\tau_1}$ and $z_2 = e^{-\Delta t/\tau_2}$ are the discrete roots, and the complex constants are defined by $\mathbf{K}_i = \frac{I_{0i} \tau_i}{2j} \left(1 - e^{-T_0/\tau_i}\right)$.

By treating the equivalent parameters $(\mathbf{K}_{\mathrm{eq}}, z_{\mathrm{eq}})$ as locally invariant over three consecutive sliding steps $\{m, m+1, m+2\}$, an equivalent dynamically updating DDC component $Z_{k,\mathrm{eq}}[m] = \mathbf{K}_{\mathrm{eq}} z_{\mathrm{eq}}^m$ approximates the combined effect of the two DDC components of $Z_k[m]$, such that:
\begin{equation} \label{eq:equiv_postulate}
    \mathbf{K}_{\mathrm{eq}} z_{\mathrm{eq}}^m = \mathbf{K}_1 z_1^m + \mathbf{K}_2 z_2^m
\end{equation}
To evaluate the instantaneous equivalent discrete root $z_{\mathrm{eq}}$, the homogeneous forward difference operator $\Delta$ is applied to \eqref{eq:equiv_postulate} to eliminate low-frequency integration errors and isolate the decay rates:
\begin{equation} \label{eq:equiv_diff}
    \mathbf{K}_{\mathrm{eq}} z_{\mathrm{eq}}^m (z_{\mathrm{eq}} - 1) = \mathbf{K}_1 z_1^m (z_1 - 1) + \mathbf{K}_2 z_2^m (z_2 - 1)
\end{equation}
By evaluating \eqref{eq:equiv_diff} at two consecutive sliding window steps $m$ and $m+1$, and taking their ratio, the equivalent characteristic root $z_{\mathrm{eq}}[m]$ is directly isolated:
\begin{equation} \label{eq:zeq_ratio_derivation}
    z_{\mathrm{eq}}[m] = \frac{\Delta Z_k[m+1]}{\Delta Z_k[m]} = \frac{\mathbf{K}_1 z_1^{m+1} (z_1 - 1) + \mathbf{K}_2 z_2^{m+1} (z_2 - 1)}{\mathbf{K}_1 z_1^m (z_1 - 1) + \mathbf{K}_2 z_2^m (z_2 - 1)}
\end{equation}
Given that the pure decay projection integral renders $\mathbf{K}_i$ purely imaginary, we define $\mathbf{K}_i = -j C_i$, where $C_i = \frac{1}{2} I_{0i} \tau_i (1 - e^{-T_0/\tau_i})$ represents the real-valued magnitude coefficient for the $i$-th mode. Substituting this into \eqref{eq:zeq_ratio_derivation}, the imaginary operators precisely cancel, yielding a purely real equivalent root:
\begin{equation} \label{eq:zeq_final_real}
    z_{\mathrm{eq}}[m] = \frac{C_1 z_1^{m+1} (1 - z_1) + C_2 z_2^{m+1} (1 - z_2)}{C_1 z_1^m (1 - z_1) + C_2 z_2^m (1 - z_2)}
\end{equation}
The result of \eqref{eq:zeq_final_real} confirms the existence of an equivalent characteristic root $z_{\mathrm{eq}}[m]$ that provides a local piecewise approximation tracking the true composite trajectory of the two DDC components. The instantaneous equivalent time constant is subsequently calculated as:
\begin{equation} \label{eq:tau_eq_dynamic}
    \tau_{\mathrm{eq}}[m] = \frac{-\Delta t}{\ln(z_{\mathrm{eq}}[m])}
\end{equation}
A similar algebraic expansion can be performed to demonstrate that this approximation holds for an arbitrary number of $n$ DDC components. Therefore, by employing the foundational single DDC homogeneous formulation ($n=1$), the proposed method can dynamically mitigate the combined effect of multiple DDC components on fundamental phasor estimation without requiring the true order $n$ a priori.

\vspace{-10mm}
\subsection{Discrete-Time Implementation for Digital Relays}
Let $T_s$ be the sampling period, $v$ the discrete sample index, and $N_s=T_0/T_s$ the number of samples per fundamental cycle. The sliding window shifts by $\Delta N = N_s / M$ samples ($\Delta t=\Delta N T_s$), where $M$ is an integer sub-cycle shift factor, making the starting sample of the $m$-th window $v_m=m\Delta N$. Modulating the sampled fault current by the $k$-th harmonic carrier yields the sliding projection over one cycle:
\begin{equation}\label{eq:discrete_sliding_sum}
\begin{aligned}
Z_k[m]=\sum_{v=m\Delta N}^{m\Delta N+N_s-1}
&x[v]\sin\left(\frac{2\pi kv}{N_s}\right)\\
&\times e^{-j2\pi kv/N_s}.
\end{aligned}
\end{equation}

For a single DDC component, its sampled form is
\begin{equation}\label{eq:discrete_ddc}
x_{\mathrm{DDC}}[v]=I_0q^v,
\end{equation}
where $q=e^{-T_s/\tau}$ is the single-sample decay factor. Applying the discrete form of Euler's identity,
\begin{equation}\label{eq:discrete_euler}
\sin\left(\frac{2\pi kv}{N_s}\right)
e^{-j2\pi kv/N_s} =
\frac{1-e^{-j4\pi kv/N_s}}{2j},
\end{equation}
the DDC contribution to the projection becomes
\begin{equation}\label{eq:discrete_split}
\begin{aligned}
Z_k[m]=&
\frac{I_0}{2j}
\sum_{v=m\Delta N}^{m\Delta N+N_s-1}q^v\\
&-\frac{I_0}{2j}
\sum_{v=m\Delta N}^{m\Delta N+N_s-1}
q^ve^{-j4\pi kv/N_s}.
\end{aligned}
\end{equation}

The first term in \eqref{eq:discrete_split} is a finite geometric series $S_d[m]$:
\begin{equation}\label{eq:discrete_decay_sum}
S_d[m] =
q^{m\Delta N}
\frac{1-q^{N_s}}{1-q}.
\end{equation}
The second term represents the discrete $2k$ oscillatory component. Being small for the selected modulation index, it is neglected. The sliding projection is thus approximated by the discrete geometric sequence:
\begin{equation}\label{eq:discrete_geom}
Z_k[m]\approx K_d z^m
\end{equation}
where the projection coefficient $K_d$ and discrete root $z$ are respectively defined as:
\begin{equation}\label{eq:Kd_definition}
K_d=
\frac{I_0}{2j}
\frac{1-q^{N_s}}{1-q}, \quad z=q^{\Delta N}=e^{-\Delta N T_s/\tau}.
\end{equation}

Although evaluating \eqref{eq:discrete_geom} directly could yield $z$, applying the forward difference $\Delta Z_k[m]=Z_k[m+1]-Z_k[m]$ provides numerical stability and filters practical projection errors. Within each active observation buffer of length $N_s + 2\Delta N$ (indexed locally by sub-windows $m \in \{0, 1, 2\}$), evaluating the differences for the first two consecutive steps gives:
\begin{equation}\label{eq:disc_dz}
\Delta Z_k[0]=K_d(z-1), \quad \Delta Z_k[1]=K_dz(z-1).
\end{equation}
Taking their ratio eliminates $K_d$ and yields the discrete decay root $z$, from which the time constant $\widehat{\tau}$ is recovered:
\begin{equation}\label{eq:disc_root}
z=\frac{\Delta Z_k[1]}{\Delta Z_k[0]}, \quad \widehat{\tau}=-\frac{\Delta N T_s}{\ln(z)}.
\end{equation}

Once $z$ is determined, $K_d$ and $q$ are obtained algebraically as
\begin{equation}\label{eq:Kd_recovery}
K_d=\frac{\Delta Z_k[0]}{z-1}, \quad q=z^{1/\Delta N}.
\end{equation}
Substituting $q$ and $K_d$ into \eqref{eq:Kd_definition} gives the estimated DDC amplitude $\widehat{I}_0$:
\begin{equation}\label{eq:I0_recovery}
\widehat{I}_0=
2jK_d
\frac{1-q}{1-q^{N_s}}.
\end{equation}
Ideally, the dominant decay projection is imaginary, making $\widehat{I}_0$ real. Small complex residuals from finite sampling are suppressed by retaining only the real-valued estimate. 

The DDC is reconstructed sample by sample using the estimated initial amplitude $\widehat{I}_0$ and time constant $\widehat{\tau}$, and is subsequently subtracted from the raw fault current $x[v]$ to produce the compensated current $x_c[v]$:
\begin{equation}\label{eq:ddc_reconstruction}
\widehat{x}_{\mathrm{DDC}}[v] =
\widehat{I}_0
e^{-vT_s/\widehat{\tau}}, \quad x_c[v] =
x[v]-\widehat{x}_{\mathrm{DDC}}[v].
\end{equation}
A conventional FCDFT is then applied to $x_c[v]$ to estimate the fundamental phasor.

For multiple DDC components tracked continuously at an arbitrary sliding step $m$, the equivalent-mode formulation retains this first-order structure. The dynamically varying equivalent root $z_{\mathrm{eq}}[m]$, time constant $\tau_{\mathrm{eq}}[m]$, and local window amplitude $\widehat{I}_{0,\mathrm{eq}}[m]$ (referenced to the start of the $m$-th window $v_m = m\Delta N$) are updated as:
\begin{equation}\label{eq:discrete_zeq}
\begin{aligned}
z_{\mathrm{eq}}[m] &=
\frac{\Delta Z_k[m+1]}
{\Delta Z_k[m]}, \quad \tau_{\mathrm{eq}}[m] =
-\frac{\Delta N T_s}
{\ln\left(z_{\mathrm{eq}}[m]\right)}, \\
\widehat{I}_{0,\mathrm{eq}}[m] &= \operatorname{Re}\!\left\{ 2j \frac{\Delta Z_k[m]}{z_{\mathrm{eq}}[m] - 1} \frac{1 - q_{\mathrm{eq}}[m]}{1 - q_{\mathrm{eq}}[m]^{N_s}} \right\},
\end{aligned}
\end{equation}
where $q_{\mathrm{eq}}[m] = \left(z_{\mathrm{eq}}[m]\right)^{1/\Delta N}$ and $\widehat{x}_{\mathrm{DDC}}[v] = \widehat{I}_{0,\mathrm{eq}}[m] e^{-(v - m\Delta N)T_s/\tau_{\mathrm{eq}}[m]}$ over the current window. Thus, the relay preserves the $n=1$ computational structure while dynamically tracking multiple decaying offsets without prior knowledge of the number of DDC components.

\vspace{-3mm}
\section{Performance Evaluation}

To validate the proposed modulation-assisted sliding-window method, a series of evaluations is conducted. The algorithm is benchmarked against the standard FCDFT and the two closest methods to the proposed approach, Method \cite{afrandideh2022} and Method \cite{7745975}, as both utilize high-frequency modulation algorithms for DDC mitigation.

The evaluation relies on three established quantitative criteria to measure the transient and steady-state performance of the phasor estimation.

The maximum overshoot ($MO$) evaluates the worst-case transient spike above the true steady-state magnitude ($X_{\mathrm{true}}$):
\begin{equation}
MO = \max_{t \ge t_f} \left( \frac{\max\!\big(X_{\mathrm{est}}(t) - X_{\mathrm{true}},\, 0\big)}{X_{\mathrm{true}}} \right) \times 100\%
\end{equation}

The rise time ($t_r$) measures the tracking speed during the initial fault transition. It represents the time required for the estimated magnitude to transition from $10\%$ to $90\%$ of the fault step change:
\begin{equation}
t_r = t_{90\%} - t_{10\%}
\end{equation}

Finally, the settling time ($t_s$) identifies when the transient effectively ends. It is defined as the elapsed time following fault inception $t_f$ required for the estimation to permanently enter and remain within a $\pm 5\%$ tolerance band around the theoretical magnitude:
\begin{equation}
t_s = \min \left\{ t > t_f \;\middle|\; \left|X_{\mathrm{est}}(u) - X_{\mathrm{true}}\right| \le 0.05 X_{\mathrm{true}},\, \forall u \ge t \right\} - t_f
\end{equation}


The modulation orders \(k=4\), 8, and 16 are considered to investigate modulation-order sensitivity. The \(k=16\) case is included for comparative purposes, although its practical applicability depends on the relay sampling rate and AAF bandwidth.



\vspace{-3mm}

\vspace{-3mm}

\subsection{Evaluation with Single DDC}

The DDC component depends on the fault inception angle and the network's equivalent $X/R$ ratio. This ratio directly dictates the time constant $\tau$ \cite{yu2021phasor}. To evaluate the robustness of the proposed methodology across a wide spectrum of fault conditions, fault currents containing single DDC are generated. The magnitude ratio ($I_0/I_1$) is varied across $\{0.1, 0.5, 1\}$, and the time constant is varied across $\tau \in \{0.5, 3, 5\}$ cycles. The proposed method is evaluated at three different modulation orders ($k \in \{4, 8, 16\}$), which yield nearly identical results for the investigated single-DDC cases.

\begin{table*}[htbp]

\centering

\scriptsize

\caption{Comparative Analysis of Algorithms for Test Signals with a Single DDC Component $k\in\{4,8,16\}$.}

\label{tab:single_ddo}

\resizebox{\textwidth}{!}{

\begin{tabular}{c c | c c c | c c c | c c c | c c c}

\hline\hline

\multicolumn{2}{c|}{\textbf{Test Condition}}

& \multicolumn{3}{c|}{\textbf{FCDFT}}

& \multicolumn{3}{c|}{\textbf{Method \cite{afrandideh2022}}}

& \multicolumn{3}{c|}{\textbf{Method \cite{7745975}}}

& \multicolumn{3}{c}{\textbf{Proposed Method}} \\

\hline

$I_0/I_1$ & $\tau$ (cyc)

& \textbf{MO (\%)} & $t_r$ \textbf{(ms)} & $t_s$ \textbf{(ms)}

& \textbf{MO (\%)} & $t_r$ \textbf{(ms)} & $t_s$ \textbf{(ms)}

& \textbf{MO (\%)} & $t_r$ \textbf{(ms)} & $t_s$ \textbf{(ms)}

& \textbf{MO (\%)} & $t_r$ \textbf{(ms)} & $t_s$ \textbf{(ms)} \\

\hline

\multirow{3}{*}{$0.1$}

& $0.5$ & 2.50 & 10.94 & 13.80 & 0.60 & 9.64 & 12.76 & 1.24 & 9.38 & 12.24 & 0.04 & 11.20 & 14.32 \\

& $3$   & 0.90 & 11.20 & 14.06 & 0.70 & 9.90 & 13.02 & 0.90 & 9.38 & 12.76 & 0.00 & 11.20 & 14.32 \\

& $5$   & 0.58 & 11.20 & 14.06 & 0.58 & 9.90 & 13.28 & 0.58 & 9.38 & 14.32 & 0.00 & 11.20 & 14.32 \\

\hline

\multirow{3}{*}{$0.5$}

& $0.5$ & 12.58 & 10.42 & 20.05 & 12.31 & 8.85 & 16.67 & 12.58 & 7.03 & 16.41 & 0.20 & 11.20 & 14.32 \\

& $3$   & 4.91 & 9.90 & 12.24 & 4.84 & 8.85 & 10.94 & 8.31 & 7.03 & 13.02 & 0.01 & 11.20 & 14.32 \\

& $5$   & 3.52 & 9.90 & 12.24 & 3.46 & 8.85 & 10.94 & 7.81 & 7.03 & 12.50 & 0.00 & 11.20 & 14.32 \\

\hline

\multirow{3}{*}{$1$}

& $0.5$ & 25.32 & 4.69 & 27.86 & 25.32 & 4.69 & 16.67 & 35.68 & 4.69 & 15.36 & 0.40 & 11.20 & 14.06 \\

& $3$   & 18.71 & 3.65 & 27.34 & 18.71 & 3.65 & 16.67 & 34.35 & 3.65 & 15.36 & 0.02 & 11.20 & 14.32 \\

& $5$   & 20.62 & 3.65 & 17.97 & 20.62 & 3.65 & 16.67 & 33.85 & 3.65 & 15.36 & 0.01 & 11.20 & 14.32 \\

\hline\hline

\end{tabular}

}

\vspace{-5mm}

\end{table*}

\begin{figure}[htbp]

\centering

\includegraphics[width=0.98\linewidth]{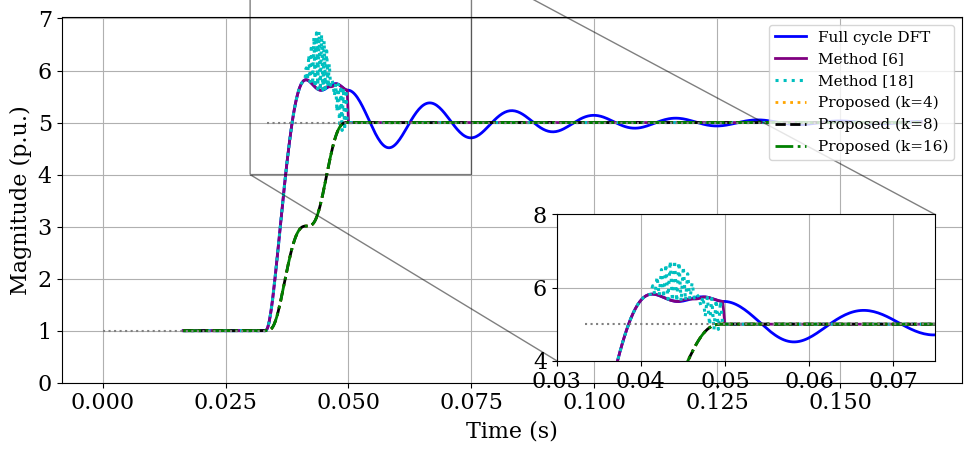}
\vspace{-4mm}
\caption{Transient magnitude estimation of fault current for a single DDC.}

\label{fig:single_ddo}

\vspace{-3mm}

\end{figure}

As observed in Table \ref{tab:single_ddo} and Fig. \ref{fig:single_ddo}, the FCDFT is highly affected by the DDC component. In case of severe faults ($I_0/I_1=1$), the FCDFT generates an overshoot up to $25.32\%$ and a settling time up to $27.86$ ms. Method \cite{7745975} exhibits larger overshoots, reaching $35.68\%$. While Method \cite{afrandideh2022} reduces the settling time in some cases, it still permits up to a $25.32\%$ overshoot.

The proposed method provides the smallest MO among the evaluated methods and maintains a settling time of $14.06$--$14.32$ ms for all investigated single-DDC cases. Regardless of the DDC parameters, the proposed method separates the isolated mode, limiting the MO to at most $0.40\%$ (and $0.00\%$--$0.02\%$ for $\tau \ge 3$ cycles). Its settling time remains tightly bounded within $14.32$ ms, indicating that the extraction accuracy is largely insensitive to the DDC time constant.

Although the transient metrics in Table~\ref{tab:single_ddo} are nearly
identical across valid modulation orders, Table~\ref{tab:k_sensitivity}
shows the effect of $k$ using higher-resolution error metrics. Increasing
$k$ reduces the retained $2k$ oscillatory residual and the resulting
estimation errors, without changing the sample-quantized rise and settling
times.

\begin{table}[!t]
\centering
\scriptsize
\caption{Modulation-Order Sensitivity for $I_0/I_1=1$, $\tau=0.5$ Cycle}
\label{tab:k_sensitivity}
\setlength{\tabcolsep}{3.5pt}
\renewcommand{\arraystretch}{1.05}
\resizebox{\columnwidth}{!}{%
\begin{tabular}{c|ccccc}
\hline
\hline
$k$ &
$\eta$ (\%) &
$E_{I_0}$ (\%) &
$E_{\max}$ (\%) &
$E_{\mathrm{RMSE}}$ (\%) &
$E_{\mathrm{MAE}}$ (\%) \\
\hline
4  & 3.9757 & 1.7022 & 0.4251 & 0.1732 & 0.1299 \\
8  & 1.9890 & 1.5864 & 0.3962 & 0.1614 & 0.1210 \\
16 & 0.9947 & 1.5624 & 0.3902 & 0.1590 & 0.1192 \\
\hline
\hline
\end{tabular}%
}
\vspace{-5mm}
\end{table}

\vspace{-5mm}
\subsection{Evaluation with two DDC}

In transmission networks, secondary currents measured by CTs often contain multiple DDC components. These originate from both the primary line properties and the internal magnetic core of the CT during saturation \cite{5109874}. A series of test signals including severe DDC conditions are evaluated. Table \ref{tab:two_ddo} details the performance across varied system time constants ($\tau_1$), secondary CT time constants ($\tau_2$), and DDC amplitude ratios ($I_{01}/I_{02}$).

\begin{figure}[htbp]

\centering

\includegraphics[width=0.98\linewidth]{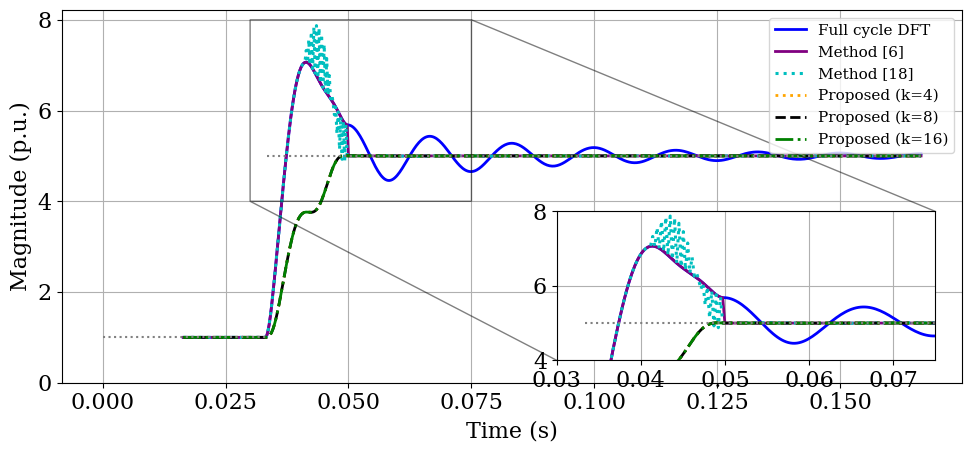}
\vspace{-4mm}
\caption{Current magnitude estimation under two DDC.}

\label{fig:dddo}

\vspace{-5mm}

\end{figure}

\begin{table*}[htbp]

\centering

\caption{Comparative Evaluations of Algorithms for Performance Under Double DC Offsets}

\label{tab:two_ddo}

\resizebox{\textwidth}{!}{

\begin{tabular}{c c c c c | c c c | c c c | c c c | c c c}

\hline\hline

\multicolumn{5}{c|}{\textbf{Test Condition}}

& \multicolumn{3}{c|}{\textbf{FCDFT}}

& \multicolumn{3}{c|}{\textbf{Method \cite{afrandideh2022}}}

& \multicolumn{3}{c|}{\textbf{Method \cite{7745975}}}

& \multicolumn{3}{c}{\textbf{Proposed Method}} \\

\hline

$I_{01}/I_1$ & $I_{01}/I_{02}$ & $\tau_1$ (cyc) & $\tau_2$ (cyc) & $k$

& \textbf{MO (\%)} & $t_r$ \textbf{(ms)} & $t_s$ \textbf{(ms)}

& \textbf{MO (\%)} & $t_r$ \textbf{(ms)} & $t_s$ \textbf{(ms)}

& \textbf{MO (\%)} & $t_r$ \textbf{(ms)} & $t_s$ \textbf{(ms)}

& \textbf{MO (\%)} & $t_r$ \textbf{(ms)} & $t_s$ \textbf{(ms)} \\

\hline

\multirow{3}{*}{$1$} & \multirow{3}{*}{$2.5$} & \multirow{3}{*}{$1$} & \multirow{3}{*}{$10$}

& $4$  & 35.20 & 3.13 & 35.94 & 35.20 & 3.13 & 16.67 & 57.58 & 3.13 & 15.36 & 0.38 & 10.94 & 13.28 \\

& & & & $8$  & 35.20 & 3.13 & 35.94 & 35.20 & 3.13 & 16.67 & 57.58 & 3.13 & 15.36 & 0.37 & 10.94 & 13.28 \\

& & & & $16$ & 35.20 & 3.13 & 35.94 & 35.20 & 3.13 & 16.67 & 57.58 & 3.13 & 15.36 & 0.37 & 10.94 & 13.28 \\

\hline

\multirow{3}{*}{$0.5$} & \multirow{3}{*}{$2.5$} & \multirow{3}{*}{$1$} & \multirow{3}{*}{$10$}

& $4$  & 10.85 & 4.95 & 19.79 & 10.84 & 4.95 & 16.67 & 19.01 & 4.95 & 15.36 & 0.09 & 11.46 & 13.80 \\

& & & & $8$  & 10.85 & 4.95 & 19.79 & 10.84 & 4.95 & 16.67 & 19.01 & 4.95 & 15.36 & 0.09 & 11.46 & 13.80 \\

& & & & $16$ & 10.85 & 4.95 & 19.79 & 10.84 & 4.95 & 16.67 & 19.01 & 4.95 & 15.36 & 0.09 & 11.46 & 13.80 \\

\hline

\multirow{3}{*}{$1$} & \multirow{3}{*}{$5$} & \multirow{3}{*}{$0.5$} & \multirow{3}{*}{$5$}

& $4$  & 26.62 & 3.65 & 27.86 & 26.61 & 3.65 & 16.67 & 45.58 & 3.65 & 15.36 & 0.62 & 11.20 & 13.80 \\

& & & & $8$  & 26.62 & 3.65 & 27.86 & 26.61 & 3.65 & 16.67 & 45.58 & 3.65 & 15.36 & 0.33 & 11.20 & 13.80 \\

& & & & $16$ & 26.62 & 3.65 & 27.86 & 26.61 & 3.65 & 16.67 & 45.58 & 3.65 & 15.36 & 0.32 & 11.20 & 13.80 \\

\hline

\multirow{3}{*}{$1$} & \multirow{3}{*}{$2$} & \multirow{3}{*}{$2$} & \multirow{3}{*}{$20$}

& $4$  & 47.91 & 2.86 & 35.94 & 47.91 & 2.86 & 16.67 & 63.14 & 2.86 & 15.36 & 0.59 & 10.68 & 13.02 \\

& & & & $8$  & 47.91 & 2.86 & 35.94 & 47.91 & 2.86 & 16.67 & 63.14 & 2.86 & 15.36 & 0.59 & 10.68 & 13.02 \\

& & & & $16$ & 47.91 & 2.86 & 35.94 & 47.91 & 2.86 & 16.67 & 63.14 & 2.86 & 15.36 & 0.59 & 10.68 & 13.02 \\

\hline\hline

\end{tabular}

}

\vspace{-5mm}

\end{table*}

Under double-DDC conditions, the FCDFT reaches large overshoots with protracted settling delays, while Method \cite{7745975} reaches a $63.14\%$ error peak under slow-decay conditions. However, the proposed method adapts to the multi-modal trajectory. As seen in Fig. \ref{fig:dddo}, it substantially reduces the compound transients, restricting the overshoot to negligible margins ($\le 0.62\%$, and $\le 0.59\%$ for $k \ge 8$) and settling within one fundamental cycle ($13.02$--$13.80$ ms).

\vspace{-3mm}

\subsection{Immunity to Measurement Noise}

Measurement signals acquired from CT secondaries inherently contain noise, typically exhibiting a signal-to-noise ratio (SNR) between $20$ to $40$ dB. To evaluate the proposed method under noise condition, Gaussian noise along with a severe fault signal ($I_0/I_1 = 1, \tau = 5$ cycles), scaling the SNR down to $20$ dB, is considered.

\begin{table*}[htbp]

\centering

\caption{Comparative Evaluations of Algorithms for Robustness under Different Noise Levels}

\label{tab:noise}

\resizebox{\textwidth}{!}{

\begin{tabular}{c c | c c c | c c c | c c c | c c c}

\hline\hline

\multicolumn{2}{c|}{\textbf{Condition}}

& \multicolumn{3}{c|}{\textbf{FCDFT}}

& \multicolumn{3}{c|}{\textbf{Method \cite{afrandideh2022}}}

& \multicolumn{3}{c|}{\textbf{Method \cite{7745975}}}

& \multicolumn{3}{c}{\textbf{Proposed Method}} \\

\hline

\textbf{SNR (dB)} & $k$

& \textbf{MO (\%)} & $t_r$ \textbf{(ms)} & $t_s$ \textbf{(ms)}

& \textbf{MO (\%)} & $t_r$ \textbf{(ms)} & $t_s$ \textbf{(ms)}

& \textbf{MO (\%)} & $t_r$ \textbf{(ms)} & $t_s$ \textbf{(ms)}

& \textbf{MO (\%)} & $t_r$ \textbf{(ms)} & $t_s$ \textbf{(ms)} \\

\hline

\multirow{3}{*}{$20$}

& $4$  & 20.64 & 3.65 & 15.62 & 20.64 & 3.65 & 17.97 & 42.35 & 3.65 & 83.85 & 0.96 & 11.72 & 15.10 \\

& $8$  & 20.64 & 3.65 & 15.62 & 20.64 & 3.65 & 17.97 & 42.35 & 3.65 & 83.85 & 0.95 & 11.72 & 14.84 \\

& $16$ & 20.64 & 3.65 & 15.62 & 20.64 & 3.65 & 17.97 & 42.35 & 3.65 & 83.85 & 1.54 & 11.72 & 14.58 \\

\hline

\multirow{3}{*}{$30$}

& $4$  & 20.63 & 3.65 & 16.41 & 20.63 & 3.65 & 16.41 & 20.63 & 3.65 & 19.53 & 0.67 & 11.46 & 14.58 \\

& $8$  & 20.63 & 3.65 & 16.41 & 20.63 & 3.65 & 16.41 & 20.63 & 3.65 & 19.53 & 0.35 & 11.20 & 14.32 \\

& $16$ & 20.63 & 3.65 & 16.41 & 20.63 & 3.65 & 16.41 & 20.63 & 3.65 & 19.53 & 2.47 & 11.46 & 14.32 \\

\hline

\multirow{3}{*}{$45$}

& $4$  & 20.62 & 3.65 & 17.97 & 20.62 & 3.65 & 16.67 & 30.84 & 3.65 & 15.36 & 0.12 & 11.20 & 14.32 \\

& $8$  & 20.62 & 3.65 & 17.97 & 20.62 & 3.65 & 16.67 & 30.84 & 3.65 & 15.36 & 0.06 & 11.20 & 14.32 \\

& $16$ & 20.62 & 3.65 & 17.97 & 20.62 & 3.65 & 16.67 & 30.84 & 3.65 & 15.36 & 0.52 & 11.20 & 14.32 \\

\hline\hline

\end{tabular}

}

\vspace{-5mm}

\end{table*}

As detailed in Table \ref{tab:noise}, the proposed method provides selective filtering due to the full-cycle trigonometric carrier integral. At $20$ dB noise, Method \cite{7745975} reaches a $42.35\%$ overshoot and an $83.85$ ms settling time, while the FCDFT and Method \cite{afrandideh2022} exhibit a $20.64\%$ overshoot. Conversely, the proposed method bounds the maximum error tightly. At $20$ dB noise, it restricts overshoot to $0.95\%$ (at $k=8$) and at most $1.54\%$ (and $\le 2.47\%$ across all evaluated noise levels) while settling within one fundamental cycle ($14.32$--$15.10$ ms). These results demonstrate that the proposed method maintains accurate phasor estimation under severe noise conditions better than other studied methods.

\vspace{-3mm}

\subsection{Interference of Harmonics}

During severe fault transients or transformer energization, the fault current contains both odd and even harmonics \cite{afrandideh2022}. A composite test signal is generated, superimposing a primary DDC ($\tau = 3$ cycles) with harmonic combinations. The first case evaluates a heavily distorted signal containing 2nd ($20\%$ of the amplitude of the fundamental harmonic) and 4th ($10\%$) harmonics. The second evaluates a spectrum containing 2nd ($20\%$), 3rd ($10\%$), and 4th ($10\%$) harmonics simultaneously.

\begin{table*}[htbp]

\centering

\caption{Comparative Performance Under Harmonic Distortion ($I_0/I_1 = 1$, $\tau = 3$ cycles)}

\label{tab:harmonics}

\resizebox{\textwidth}{!}{

\begin{tabular}{c c | c c c | c c c | c c c | c c c}

\hline\hline

\multicolumn{2}{c|}{\textbf{Harmonic Content}}

& \multicolumn{3}{c|}{\textbf{FCDFT}}

& \multicolumn{3}{c|}{\textbf{Method \cite{afrandideh2022}}}

& \multicolumn{3}{c|}{\textbf{Method \cite{7745975}}}

& \multicolumn{3}{c}{\textbf{Proposed Method}} \\

\hline

\textbf{Distortion} & $k$

& \textbf{MO (\%)} & $t_r$ \textbf{(ms)} & $t_s$ \textbf{(ms)}

& \textbf{MO (\%)} & $t_r$ \textbf{(ms)} & $t_s$ \textbf{(ms)}

& \textbf{MO (\%)} & $t_r$ \textbf{(ms)} & $t_s$ \textbf{(ms)}

& \textbf{MO (\%)} & $t_r$ \textbf{(ms)} & $t_s$ \textbf{(ms)} \\

\hline

\multirow{3}{*}{2nd (20\%) \& 4th (10\%)}

& $4$  & 19.79 & 3.39 & 27.34 & 19.79 & 3.39 & 16.67 & 26.35 & 3.39 & 16.41 & 0.02 & 11.98 & 14.58 \\

& $8$  & 19.79 & 3.39 & 27.34 & 19.79 & 3.39 & 16.67 & 26.35 & 3.39 & 16.41 & 0.02 & 11.98 & 14.58 \\

& $16$ & 19.79 & 3.39 & 27.34 & 19.79 & 3.39 & 16.67 & 26.35 & 3.39 & 16.41 & 0.02 & 11.98 & 14.58 \\

\hline

\multirow{3}{*}{2nd (20\%), 3rd (10\%), 4th (10\%)}

& $4$  & 19.79 & 3.12 & 27.34 & 19.79 & 3.12 & 16.67 & 27.17 & 3.12 & 16.41 & 0.02 & 12.24 & 14.58 \\

& $8$  & 19.79 & 3.12 & 27.34 & 19.79 & 3.12 & 16.67 & 27.17 & 3.12 & 16.41 & 0.02 & 12.24 & 14.58 \\

& $16$ & 19.79 & 3.12 & 27.34 & 19.79 & 3.12 & 16.67 & 27.17 & 3.12 & 16.41 & 0.02 & 12.24 & 14.58 \\

\hline\hline

\end{tabular}

}

\vspace{-5mm}

\end{table*}

Under the proposed methodology, orthogonal sinusoids integrate to exactly zero over the fundamental window $T_0$. By selecting an even carrier parameter (e.g., $k=4, 8, 16$), the algorithm rejects the investigated low-frequency harmonic components. As Table \ref{tab:harmonics} shows, these prominent fault harmonics contribute strictly zero energy to the sliding projection integrals. Thus, the proposed method achieves a negligible $0.02\%$ MO (matching the harmonic-free baseline) with a $14.58$ ms settling time and isolates the DDC component regardless of harmonic severity.

\vspace{-5mm}

\subsection{Impact of Sampling Rate}

To guarantee compatibility with a wide spectrum of commercial digital relays, the proposed method must remain stable across different sampling resolutions. The algorithm is evaluated at standardized sampling rates extending from a sparse $960$ Hz ($N=16$ samples/cycle) up to $7680$ Hz ($N=128$) \cite{10158378}.

\begin{table*}[htbp]

\centering

\caption{Impact of Sampling Rate on Single DDC Estimation ($I_0/I_1=1.0$, $\tau=3.0$ cyc)}

\label{tab:sampling}

\resizebox{\textwidth}{!}{

\begin{tabular}{c c | c c c | c c c | c c c | c c c}

\hline\hline

\multicolumn{2}{c|}{\textbf{Sampling}}

& \multicolumn{3}{c|}{\textbf{FCDFT}}

& \multicolumn{3}{c|}{\textbf{Method \cite{afrandideh2022}}}

& \multicolumn{3}{c|}{\textbf{Method \cite{7745975}}}

& \multicolumn{3}{c}{\textbf{Proposed Method}} \\

\hline

$F_s$ \textbf{(Hz)} & $k$

& \textbf{MO (\%)} & $t_r$ \textbf{(ms)} & $t_s$ \textbf{(ms)}

& \textbf{MO (\%)} & $t_r$ \textbf{(ms)} & $t_s$ \textbf{(ms)}

& \textbf{MO (\%)} & $t_r$ \textbf{(ms)} & $t_s$ \textbf{(ms)}

& \textbf{MO (\%)} & $t_r$ \textbf{(ms)} & $t_s$ \textbf{(ms)} \\

\hline

\multirow{3}{*}{$960\;(N=16)$}

& $4$  & 18.77 & 3.12 & 19.79 & 18.77 & 3.12 & 16.67 & 29.27 & 3.12 & 15.62 & 0.09 & 11.46 & 14.58 \\

& $8$  & -- & -- & -- & -- & -- & -- & -- & -- & -- & -- & -- & -- \\

& $16$ & -- & -- & -- & -- & -- & -- & -- & -- & -- & -- & -- & -- \\

\hline

\multirow{3}{*}{$1920\;(N=32)$}

& $4$  & 18.75 & 3.65 & 19.27 & 18.75 & 3.65 & 16.67 & 32.59 & 3.65 & 14.06 & 0.05 & 10.94 & 14.06 \\

& $8$  & 18.75 & 3.65 & 19.27 & 18.75 & 3.65 & 16.67 & 32.59 & 3.65 & 14.06 & 0.05 & 10.94 & 14.06 \\

& $16$ & -- & -- & -- & -- & -- & -- & -- & -- & -- & -- & -- & -- \\

\hline

\multirow{3}{*}{$3840\;(N=64)$}

& $4$  & 18.71 & 3.65 & 27.34 & 18.71 & 3.65 & 16.67 & 34.35 & 3.65 & 15.36 & 0.02 & 11.20 & 14.32 \\

& $8$  & 18.71 & 3.65 & 27.34 & 18.71 & 3.65 & 16.67 & 34.35 & 3.65 & 15.36 & 0.02 & 11.20 & 14.32 \\

& $16$ & 18.71 & 3.65 & 27.34 & 18.71 & 3.65 & 16.67 & 34.35 & 3.65 & 15.36 & 0.02 & 11.20 & 14.32 \\

\hline

\multirow{3}{*}{$7680\;(N=128)$}

& $4$  & 18.69 & 3.65 & 27.21 & 18.69 & 3.65 & 16.67 & 35.12 & 3.65 & 16.02 & 0.01 & 11.20 & 14.32 \\

& $8$  & 18.69 & 3.65 & 27.21 & 18.69 & 3.65 & 16.67 & 35.12 & 3.65 & 16.02 & 0.01 & 11.20 & 14.32 \\

& $16$ & 18.69 & 3.65 & 27.21 & 18.69 & 3.65 & 16.67 & 35.12 & 3.65 & 16.02 & 0.01 & 11.20 & 14.32 \\

\hline\hline

\end{tabular}

}

\vspace{-5mm}

\end{table*}

The numerical results in Table \ref{tab:sampling} confirm that the discrete characteristic root extraction remains well conditioned at all practical sampling frequencies, provided the modulation order obeys the Nyquist limit ($k < N/2$). When the Nyquist condition is violated, the compensation becomes invalid (e.g., $k \ge 8$ at $960$ Hz). However, as long as the Nyquist condition is met (e.g., $k=4$ at $960$ Hz, or $k \le 16$ at $3840$ Hz), the proposed method maintains a negligible overshoot ($0.01\%$--$0.09\%$) and fast settling ($14.06$--$14.58$ ms) with accurate estimation.

\vspace{-3mm}

\subsection{Sensitivity to Off-Nominal Frequency}

\begin{figure}[htbp]

\centering

\includegraphics[width=0.98\linewidth]{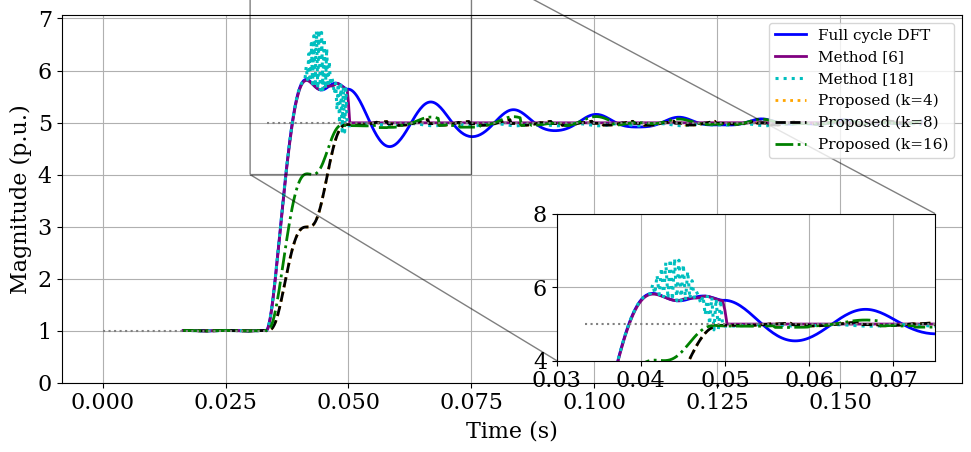}
\vspace{-4mm}
\caption{Impact of a static off-nominal frequency drop ($59.5$ Hz) on estimation accuracy.}

\label{fig:off_nominal_low}

\vspace{-5mm}

\end{figure}

\begin{figure}[htbp]

\centering

\includegraphics[width=0.98\linewidth]{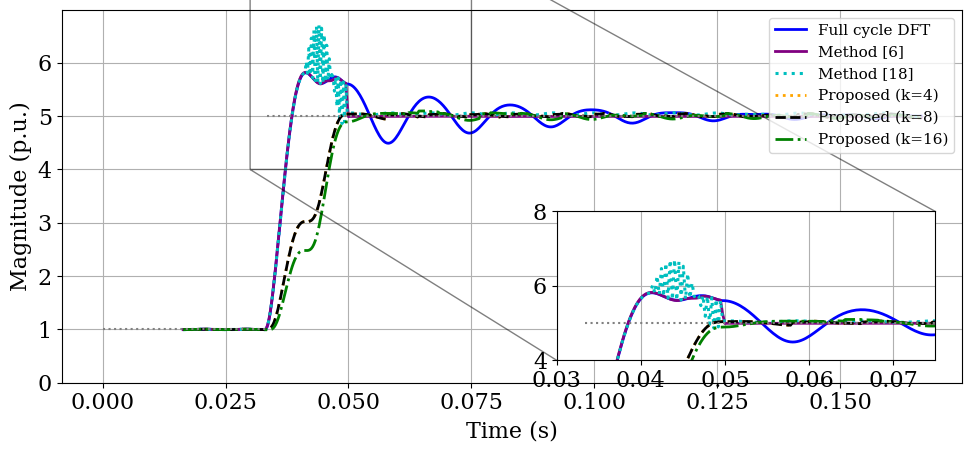}
\vspace{-4mm}
\caption{Impact of a static off-nominal frequency rise ($60.5$ Hz) on estimation accuracy.}

\label{fig:off_nominal_high}

\vspace{-5mm}

\end{figure}

\begin{figure}[htbp]

\centering

\includegraphics[width=0.98\linewidth]{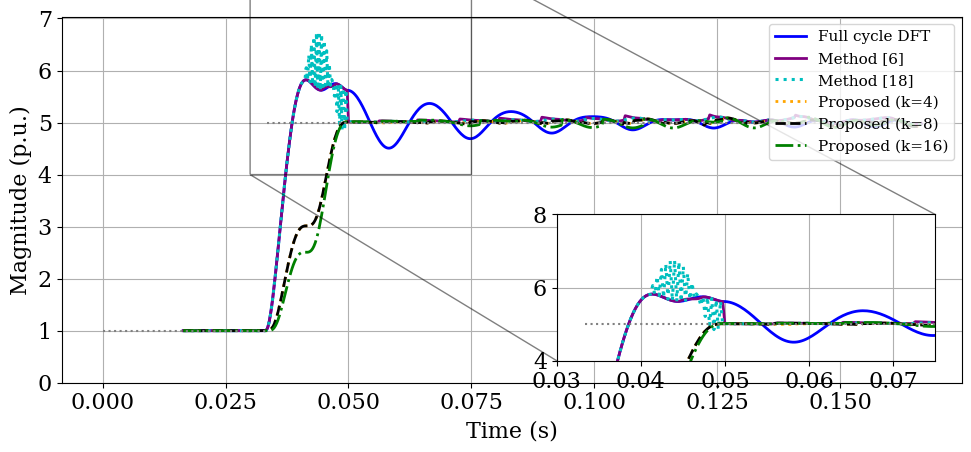}
\vspace{-3mm}
\caption{Impact of dynamic frequency variation (3\% deviation) on estimation accuracy.}

\label{fig:dynamic_freq}

\vspace{-5mm}

\end{figure}

During grid disturbances, such as multi-mode power swings or large load switching events, the system frequency can deviate from its nominal value. The frequency deviation is usually allowed in the range of 2-3\% for a short period \cite{luo2006estimation}. To evaluate this, the estimators were tested against static off-nominal frequencies ($59.5$ Hz and $60.5$ Hz in Figs. \ref{fig:off_nominal_low} and \ref{fig:off_nominal_high}) as well as a dynamic condition with continuous $3\%$ linear frequency fluctuation (see Fig. \ref{fig:dynamic_freq}).

As shown in the figures \ref{fig:off_nominal_low} to \ref{fig:dynamic_freq}, the proposed method is not sensitive to normal frequency variations in the grid and is able to reconstruct the fundamental harmonic even better than benchmarked methods.

\begin{table}[!t]
\centering
\caption{Comparison of the Computation Burden and Number of Samples Required for Phasor Estimation Algorithms}
\label{tab:complexity}
\tiny
\setlength{\tabcolsep}{3pt}
\begin{tabular}{@{}l c c c c c@{}}
\toprule
Algorithm &
\shortstack{DDC extraction \\ operation} &
DFT &
\shortstack{Additional \\ nonlinear ops.} &
\shortstack{Required number \\ of samples} &
Complexity \\
\midrule
FCDFT
& -- 
& $N$-point 
& -- 
& $N$ 
& $\mathcal{O}(N)$ \\

Method \cite{afrandideh2022}
& \shortstack{Two $N$-sample averaging \\ filters, three outputs}
& $N$-point
& Square root
& $N+2$
& $\mathcal{O}(N)$ \\

Method \cite{7745975}
& \shortstack{Two $N$-sample sums \\ (original and modulated)}
& $N$-point
& --
& $N$
& $\mathcal{O}(N)$ \\

Proposed Algorithm
& \shortstack{Three modulated $N$-sample \\ sliding projections}
& $N$-point
& Log/exp
& $N+2N/M$
& $\mathcal{O}(N)$ \\
\bottomrule
\end{tabular}
\vspace{-5mm}
\end{table}


\vspace{-3mm}
\subsection{Computational Complexity}

The computational complexity and required number of samples of the evaluated phasor estimation algorithms are summarized in Table \ref{tab:complexity}. All estimators employ an $N$-point DFT for fundamental phasor estimation. While the standard FCDFT requires no DDC compensation, Method \cite{afrandideh2022}, Method \cite{7745975}, and the proposed method additionally perform high-frequency modulation-based processing to estimate or compensate the DDC component. Method \cite{afrandideh2022} combines three successive outputs of one-cycle averaging filters applied to the original and modulated signals and therefore requires $N+2$ samples. Method \cite{7745975} obtains the DDC information from $N$-sample summations of the original and high-frequency-modulated signals and requires only one cycle ($N$ samples).

The proposed method uses three modulation-assisted sliding projections separated by $\Delta N = N/M$ samples, spanning an observation buffer of $N+2N/M$ samples for parameter extraction (e.g., $80$ samples or $1.25$ fundamental cycles for $N=64$ and $M=8$). Once extracted, the DDC parameters are used to reconstruct and subtract the decaying offset across the active $N$-sample DFT window, enabling the compensated phasor magnitude to enter the 5\% settling band within one fundamental cycle ($\approx 14\text{ ms}$). Furthermore, because parameter updates occur every $N/M$ samples and reuse previously computed sliding projections, only one new projection is evaluated per sub-cycle shift. Although the parameter extraction and time-domain reconstruction involve logarithmic and exponential operations (whereas Method \cite{afrandideh2022} employs an algebraic square-root calculation), all three DDC-compensated algorithms retain linear $\mathcal{O}(N)$ computational complexity because their dominant operations consist of finite-length summations, modulation, and DFT processing. Thus, the additional computational burden of the proposed method remains compatible with real-time implementation in digital protective relays \cite{phadke2009computer}.

\vspace{-3mm}
\subsection{Simulation-Based Analysis}

\begin{figure}[htbp]
    \centering
    \includegraphics[width=0.9\linewidth]{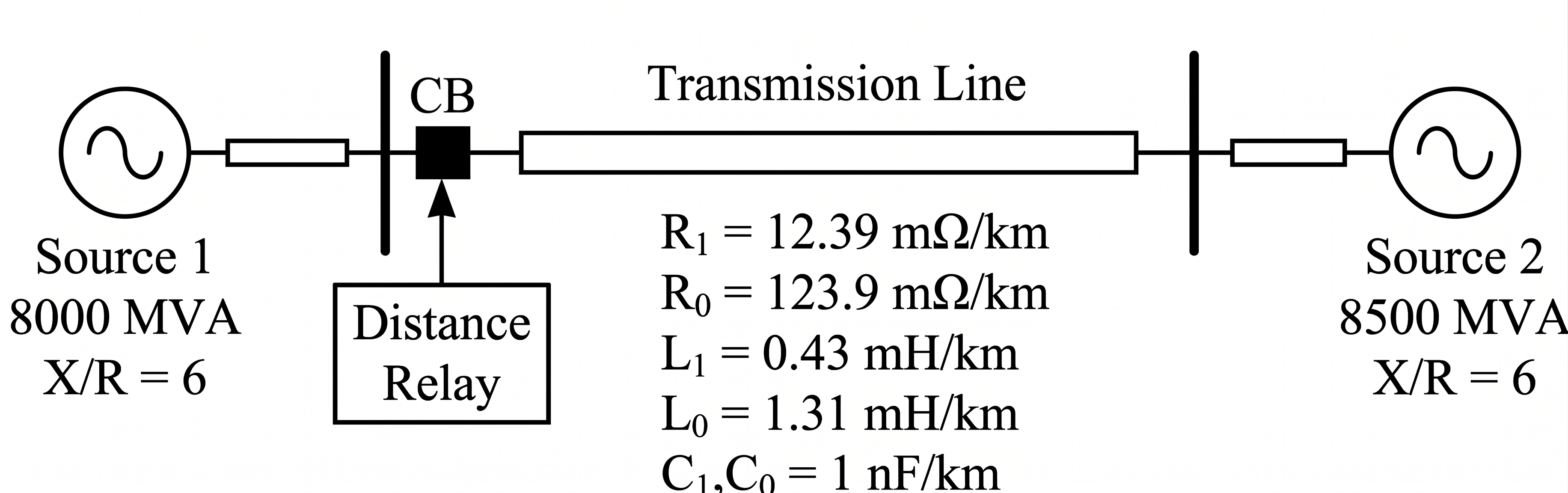}
    \vspace{-3mm}
    \caption{Single-line diagram of the tested grid adapted from \cite{hajimohammadi2024}}
    \label{test system}
    \vspace{-3mm}
\end{figure}

The dynamic performance of the phasor-estimation methods is coded and evaluated using a Python-based model of a $230$~kV, $60$~Hz, $100$~km transmission line supplied from two equivalent sources. The source impedances are derived from short-circuit capacities of $8000$~MVA and $8500$~MVA with an $X/R$ ratio of 6 (see Fig. \ref{test system} for more details). The transmission line is represented using its positive- and zero-sequence series parameters. $V_{base}$ = 230 kV and $S_{base}$ = 800 MVA are selected for per unit conversion. A solid single-phase-to-ground (A-G) fault with zero fault resistance is applied at $t_f=200$~ms.

Two complementary evaluations are performed. First, a fault at $50\%$ of the line is used to assess the transient magnitude-estimation performance of the different methods. Second, faults at $77\%$ and $85\%$ of the line are used to investigate their effect on the apparent-impedance trajectories of the distance relay. The relay is located at the Source-1 terminal, and its Zone~1 Mho characteristic is configured to cover $80\%$ of the protected line. Thus, the $77\%$ fault represents a boundary-adjacent internal fault, whereas the $85\%$ fault represents a nearby external fault.

\begin{figure}[htbp]
\centering
\includegraphics[width=0.98\linewidth]{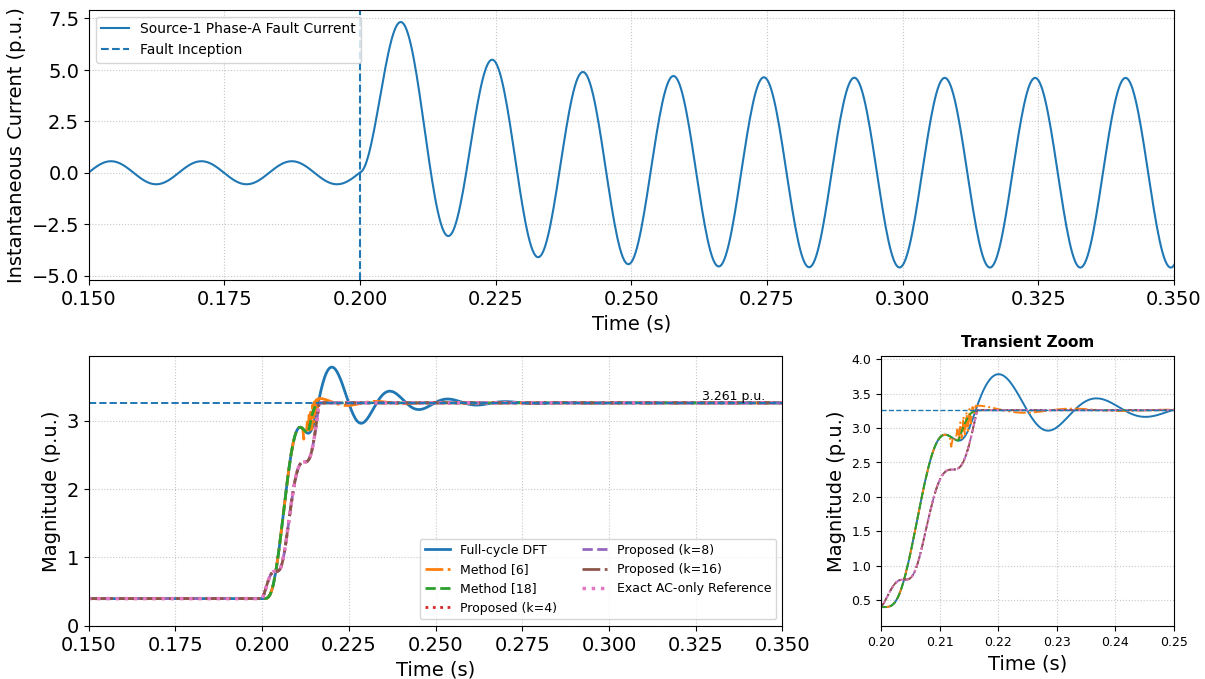}
\vspace{-4mm}
\caption{Estimated fundamental-current magnitude for the A-G fault at $50\%$ of the protected line.}
\label{fig:phasor_50}
\vspace{-3mm}
\end{figure}

The $50\%$ fault is used to quantify the transient tracking performance in terms of MO, rise time, and settling time. Table~\ref{tab:dynamic_performance} summarizes the corresponding results for the FCDFT, Method~\cite{afrandideh2022}, Method~\cite{7745975}, and the proposed method with modulation orders $k=4$, $8$, and $16$.



\begin{table}[t]
\centering
\scriptsize
\caption{Transient performance for the A-G fault at $50\%$ of the line.}
\label{tab:dynamic_performance}
\begin{tabular}{lccc}
\hline
Method & $MO$ (\%) & $t_r$ (ms) & $t_s$ (ms) \\
\hline
Full-cycle DFT                         & 16.041 & 11.458 & 37.760 \\
Method \cite{afrandideh2022}           &  2.016 &  9.375 & 15.625 \\
Method \cite{7745975}                  &  0.004 & 10.417 & 14.583 \\
Proposed ($k \in \{4,8,16\}$)          &  0.001 & 13.802 & 16.146 \\
\hline
\end{tabular}
\vspace{-5mm}
\end{table}

As shown in Fig.~\ref{fig:phasor_50} and Table~\ref{tab:dynamic_performance}, the conventional FCDFT is strongly affected by the DDC component and exhibits the largest transient deviation, with a MO of $16.041\%$ and a settling time of $37.760$~ms. Method~\cite{afrandideh2022} reduces the MO to $2.016\%$ and provides the shortest rise time of $9.375$~ms, with a settling time of $15.625$~ms. Method~\cite{7745975} further limits the overshoot to $0.004\%$ and provides the shortest settling time of $14.583$~ms. The proposed method provides the smallest MO of $0.001\%$ for all three modulation orders. Its settling time is $16.146$~ms, which is slightly less than one fundamental cycle at $60$~Hz. The identical results obtained for $k=4$, $8$, and $16$ indicate that, for the considered fault waveform, the modulation order has negligible influence on the transient-magnitude estimation performance.

The influence of phasor-estimation errors on distance-relay operation is examined separately using faults close to the Zone~1 boundary. The $77\%$ fault lies inside the $80\%$ Zone~1 reach and is therefore used to assess dependability near the protection boundary. In contrast, the $85\%$ fault lies immediately outside Zone~1 and is used to evaluate relay response against transient overreach.

\begin{figure}[htbp]
\centering
\includegraphics[width=0.95\linewidth]{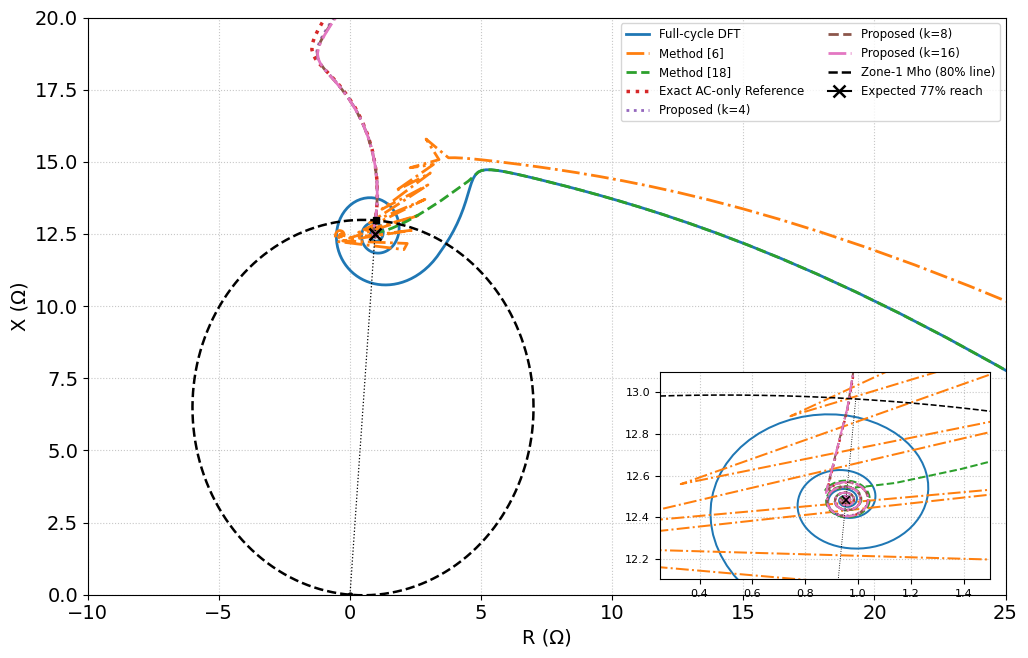}
\vspace{-4mm}
\caption{Apparent-impedance trajectories for the internal A-G fault at $77\%$ of the protected line, close to the Zone~1 Mho boundary.}
\label{fig:impedance_77}
\vspace{-5mm}
\end{figure}

For the $77\%$ internal fault, the theoretical fault impedance lies close to the Zone~1 characteristic, making the relay decision particularly sensitive to transient phasor-estimation errors. As illustrated in Fig.~\ref{fig:impedance_77}, the uncompensated FCDFT exhibits a relatively large transient loop around the expected fault impedance before convergence. Method~\cite{afrandideh2022} also produces noticeable transient excursions, whereas Method~\cite{7745975} approaches the expected operating point with reduced local fluctuation. In comparison, the proposed method produces the most confined trajectories, with the results for $k=4$, $8$, and $16$ closely overlapping around the expected $77\%$ reach point. This behavior is particularly important for a boundary-adjacent internal fault, where excessive transient displacement can delay or destabilize the in-zone decision.

\begin{figure}[htbp]
\centering
\includegraphics[width=0.95\linewidth]{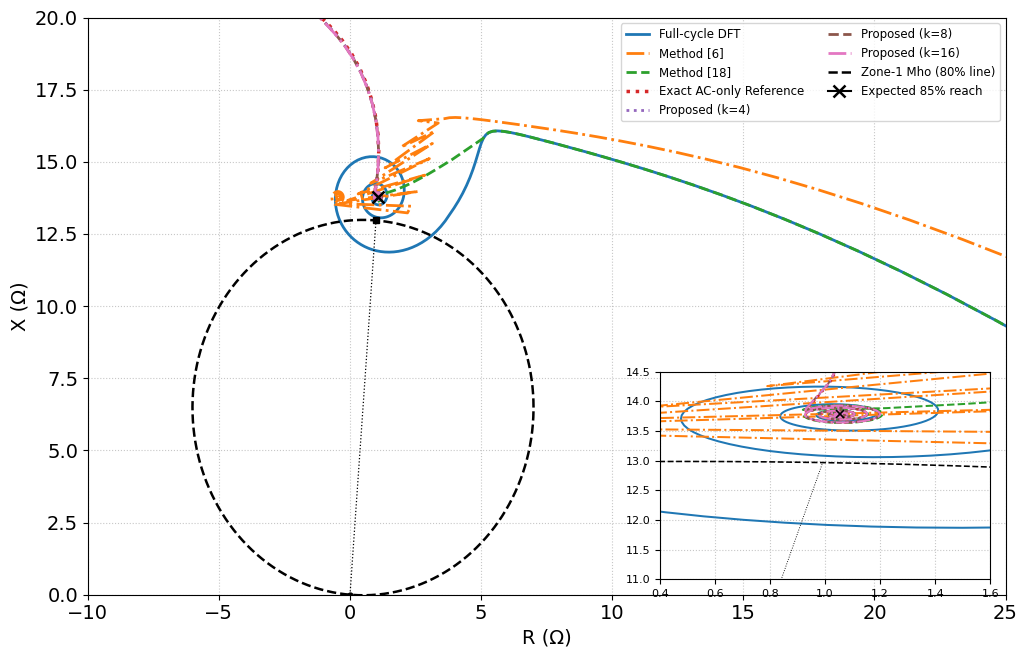}
\vspace{-4mm}
\caption{Apparent-impedance trajectories for the external A-G fault at $85\%$ of the protected line, outside the Zone~1 Mho boundary.}
\label{fig:impedance_85}
\vspace{-5mm}
\end{figure}

The $85\%$ fault provides a complementary reliability test because its theoretical apparent impedance lies outside Zone~1. As shown in Fig.~\ref{fig:impedance_85}, transient errors in the estimated current phasor displace the apparent-impedance trajectories during the initial fault period. The FCDFT exhibits a large transient loop extending toward and into the Zone~1 characteristic, indicating susceptibility to transient overreach. Method~\cite{afrandideh2022} and Method~\cite{7745975} reduce the local oscillation around the final operating point, although noticeable transient excursions remain. In contrast, the proposed method maintains tightly confined trajectories around the expected $85\%$ reach point for all investigated modulation orders and remains outside the Zone~1 characteristic. This reduced transient displacement decreases the risk of transient overreach and improves Zone~1 discrimination for the investigated external fault.

\vspace{-3mm}
\section{Conclusion}
This paper proposed a modulation-assisted algorithm for fault current phasor estimation that effectively eliminates DDC offsets causing severe spectral leakage in conventional DFT. By modulating the signal with a high-order harmonic carrier, the algorithm isolates DDC offsets and utilizes a forward difference operator to dynamically track multiple DDC components as a single equivalent decaying signal. Extensive evaluations validated the algorithm's accuracy, demonstrating a MO strictly bounded below 0.62\% under complex double-DDC conditions and settling within one fundamental cycle (under 13.80 ms). Furthermore, the method maintained robust performance under severe 20 dB noise, limiting the MO to 2.47\%. Ultimately, this approach guarantees fast, secure distance relay operations during complex Zone 1 grid faults.

\vspace{-5mm}
\section*{Acknowledgments}
The authors acknowledge the use of AI tools for English writing assistance, grammar checking, and mathematical derivation verification in the preparation of this manuscript.

\vspace{-5mm}
\bibliographystyle{IEEEtran}
\bibliography{ref}

@article{benmouyal1995,
  author  = {Benmouyal, G.},
  title   = {Removal of {DC}-offset in current waveforms using digital mimic filtering},
  journal = {IEEE Transactions on Power Delivery},
  volume  = {10},
  number  = {2},
  pages   = {621--630},
  year    = {1995},
  doi     = {10.1109/61.400869}
}

@article{guo2003,
  author  = {Guo, Y. and Kezunovic, M. and Chen, D.},
  title   = {Simplified algorithms for removal of the effect of exponentially decaying {DC}-offset on the Fourier algorithm},
  journal = {IEEE Transactions on Power Delivery},
  volume  = {18},
  number  = {3},
  pages   = {711--717},
  year    = {2003},
  doi     = {10.1109/TPWRD.2003.813894}
}

@article{kang2009,
  author  = {Kang, S. H. and Lee, D.-G. and Nam, S.-R. and Crossley, P. A. and Kang, Y.-C.},
  title   = {Fourier transform-based modified phasor estimation method immune to the effect of the {DC} offsets},
  journal = {IEEE Transactions on Power Delivery},
  volume  = {24},
  number  = {3},
  pages   = {1104--1111},
  year    = {2009},
  doi     = {10.1109/TPWRD.2009.2014032}
}

@article{altallaq2011,
  author  = {Al-Tallaq, K. N. A. and Putrus, G. A. and Li, X.},
  title   = {Online algorithm for removal of decaying {DC}-offset from phasor estimates},
  journal = {Electric Power Systems Research},
  volume  = {81},
  number  = {12},
  pages   = {2209--2218},
  year    = {2011}
}

@inproceedings{ren2009,
  author    = {Ren, J. and Kezunovic, M.},
  title     = {Elimination of {DC} offset in accurate phasor estimation using recursive wavelet transform},
  booktitle = {2009 IEEE Bucharest PowerTech},
  pages     = {1--5},
  year      = {2009}
}

@article{abdoos2016,
  author  = {Abdoos, A. A. and Ghataki, M. K. and Sadeh, J.},
  title   = {Accurate and fast {DC} offset removal method for digital relaying schemes},
  journal = {IET Generation, Transmission \& Distribution},
  volume  = {10},
  number  = {8},
  pages   = {1877--1885},
  year    = {2016},
  doi     = {10.1049/iet-gtd.2014.1150}
}

@article{mohammadi2022review,
  author  = {Mohammadi, S. and Mahmoudi, A. and Kahourzade, S. and Yazdani, A. and Shafiullah, G. M.},
  title   = {Decaying {DC} offset current mitigation in phasor estimation applications: A review},
  journal = {Energies},
  volume  = {15},
  number  = {14},
  pages   = {5260},
  year    = {2022},
  doi     = {10.3390/en15145260}
}

@article{afrandideh2022,
  author  = {Afrandideh, S. and Haghjoo, F.},
  title   = {A {DFT}-based phasor estimation method robust to primary and secondary decaying {DC} components},
  journal = {Electric Power Systems Research},
  volume  = {208},
  pages   = {107907},
  year    = {2022},
  doi     = {10.1016/j.epsr.2022.107907}
}

@article{mohammadi2022novel,
  title={A novel analytical method for DC offset mitigation enhancing DFT phasor estimation},
  author={Mohammadi, Sina and Rezaei, Navid and Mahmoudi, Amin and others},
  journal={Electric Power Systems Research},
  volume={209},
  pages={108036},
  year={2022},
  publisher={Elsevier}
}

@ARTICLE{7745975,
  author={Jafarpisheh, Babak and Madani, Seyed M. and Mohammad Shahrtash, S.},
  journal={IEEE Transactions on Power Delivery}, 
  title={A New DFT-Based Phasor Estimation Algorithm Using High-Frequency Modulation}, 
  year={2017},
  volume={32},
  number={6},
  pages={2416-2423},
  doi={10.1109/TPWRD.2016.2629762}}

@article{hajimohammadi2024,
  author    = {Haji-Mohammadi, Amir and Sanaye-Pasand, Majid},
  title     = {Novel Method for Decaying DC Component Removal Based on Solving the Differential Equations},
  journal   = {IEEE Transactions on Power Delivery},
  year      = {2024},
  publisher = {IEEE}
}

@article{jamali2022,
  author    = {Jamali, Sadegh and Zamani, Hamid},
  title     = {Removal of System and CT Decaying DC Component in Fault Signal Using Selective Interval Integration},
  journal   = {IEEE Transactions on Power Delivery},
  volume    = {37},
  number    = {2},
  pages     = {905--912},
  year      = {2022},
  publisher = {IEEE},
  doi       = {10.1109/TPWRD.2021.3075303}
}

@article{parpaei2022,
  author    = {Parpaei, Mohammad and Askarian Abyaneh, Hossein and Razavi, Farzad},
  title     = {An innovative method to eliminate multiple exponentially decaying DC components based on the discrete Fourier transform in the numerical distance relay},
  journal   = {IET Generation, Transmission \& Distribution},
  volume    = {16},
  number    = {18},
  pages     = {3731--3743},
  year      = {2022},
  publisher = {Wiley/IET},
  doi       = {10.1049/gtd2.12408}
}

@article{yu2006discrete,
  title={A discrete Fourier transform-based adaptive mimic phasor estimator for distance relaying applications},
  author={Yu, Chi-Shan},
  journal={IEEE Transactions on Power delivery},
  volume={21},
  number={4},
  pages={1836--1846},
  year={2006},
  publisher={IEEE}
}

@article{jiang2014modified,
  title={A modified empirical mode decomposition filtering-based adaptive phasor estimation algorithm for removal of exponentially decaying DC offset},
  author={Jiang, Zhen and Miao, Shihong and Liu, Pei},
  journal={IEEE transactions on power delivery},
  volume={29},
  number={3},
  pages={1326--1334},
  year={2014},
  publisher={IEEE}
}

@article{pazoki2017new,
  title={A new DC-offset removal method for distance-relaying application using intrinsic time-scale decomposition},
  author={Pazoki, Mohammad},
  journal={IEEE Transactions on Power Delivery},
  volume={33},
  number={2},
  pages={971--980},
  year={2017},
  publisher={IEEE}
}

@article{jafarpisheh2018phasor,
  title={Phasor estimation algorithm based on complex frequency filters for digital relaying},
  author={Jafarpisheh, Babak and Madani, Seyed M and Parvaresh, Farzad},
  journal={IEEE Transactions on Instrumentation and Measurement},
  volume={67},
  number={3},
  pages={582--592},
  year={2018},
  publisher={IEEE}
}

@article{silva2017modified,
  title={Modified DFT-based phasor estimation algorithms for numerical relaying applications},
  author={Silva, Kleber Melo and Nascimento, Francisco AO},
  journal={IEEE transactions on power delivery},
  volume={33},
  number={3},
  pages={1165--1173},
  year={2017},
  publisher={IEEE}
}

@article{hajizadeh2020analytical,
  title={An analytical fast decaying DC mitigation method for digital relaying applications},
  author={Hajizadeh, Hamid and Ahmadi, Seyed-Alireza and Sanaye-Pasand, Majid},
  journal={IEEE Transactions on Power Delivery},
  volume={36},
  number={6},
  pages={3529--3537},
  year={2020},
  publisher={IEEE}
}

@article{kumar2021novel,
  title={A novel sub-cycle-based method for estimation of decaying DC component and fundamental phasor},
  author={Kumar, Bandi Ravi and Mohapatra, Abheejeet and Chakrabarti, Saikat},
  journal={IEEE Transactions on Instrumentation and Measurement},
  volume={70},
  pages={1--10},
  year={2021},
  publisher={IEEE}
}

@article{jafarian2011weighted,
  title={Weighted least error squares based variable window phasor estimator for distance relaying application},
  author={Jafarian, P and Sanaye-Pasand, M},
  journal={IET generation, transmission \& distribution},
  volume={5},
  number={3},
  pages={298--306},
  year={2011},
  publisher={IET}
}

@article{hwang2018least,
  title={Least error squared phasor estimation with identification of a decaying DC component},
  author={Hwang, Jin Kwon},
  journal={IET Generation, Transmission \& Distribution},
  volume={12},
  number={7},
  pages={1486--1492},
  year={2018},
  publisher={Wiley Online Library}
}

@article{mahari2017adaptive,
  title={Adaptive phasor estimation algorithm to enhance numerical distance protection},
  author={Mahari, Arash and Sanaye-Pasand, Majid and Hashemi, Sayyed Mohammad},
  journal={IET Generation, Transmission \& Distribution},
  volume={11},
  number={5},
  pages={1170--1178},
  year={2017},
  publisher={Wiley Online Library}
}

@article{tajdinian2016analytical,
  title={An analytical approach for removal of decaying DC component considering frequency deviation},
  author={Tajdinian, Mohsen and Jahromi, Mehdi Zareian and Mohseni, Kazem and Kouhsari, Shahram Montaser},
  journal={Electric Power Systems Research},
  volume={130},
  pages={208--219},
  year={2016},
  publisher={Elsevier}
}

@article{achlerkar2019assessment,
  title={Assessment of DC offset in fault current signal for accurate phasor estimation considering current transformer response},
  author={Achlerkar, Pankaj D and Panigrahi, Bijaya Ketan},
  journal={IET Science, Measurement \& Technology},
  volume={13},
  number={3},
  pages={403--408},
  year={2019},
  publisher={Wiley Online Library}
}

@article{hwang2018dft,
  title={DFT-based phasor estimation for removal of the effect of multiple DC components},
  author={Hwang, Jin Kwon and Song, Chul Ki and Jeong, Moon Gyu},
  journal={IEEE Transactions on Power Delivery},
  volume={33},
  number={6},
  pages={2901--2909},
  year={2018},
  publisher={IEEE}
}

@article{tajdinian2017half,
  title={Half-cycle method for exponentially DC components elimination applicable in phasor estimation},
  author={Tajdinian, Mohsen and Seifi, Ali Reza and Allahbakhshi, Mehdi},
  journal={IET Science, Measurement \& Technology},
  volume={11},
  number={8},
  pages={1032--1042},
  year={2017},
  publisher={Wiley Online Library}
}

@article{allahbakhshi2019hybrid,
  title={Hybrid approach for immunisation of DFT-based phasor estimation method against decaying DC components},
  author={Allahbakhshi, Mehdi and Tajdinian, Mohsen and Seifi, Ali Reza and Zareian Jahromi, Mehdi and Bagheri, Alireza},
  journal={IET Science, Measurement \& Technology},
  volume={13},
  number={2},
  pages={238--246},
  year={2019},
  publisher={Wiley Online Library}
}

@article{yu2021phasor,
  title={A phasor estimation algorithm robust to decaying DC component},
  author={Yu, Haotian and Jin, Zongshuai and Zhang, Hengxu and Terzija, Vladimir},
  journal={IEEE Transactions on Power Delivery},
  volume={37},
  number={2},
  pages={860--870},
  year={2021},
  publisher={IEEE}
}

@inproceedings{luo2006estimation,
  title={Estimation of wind penetration as limited by frequency deviation},
  author={Luo, Changling and Far, H Golestani and Banakar, Hadi and Keung, Pin-Kwan and Ooi, Boon-Teck},
  booktitle={2006 IEEE Power Engineering Society General Meeting},
  pages={8--pp},
  year={2006},
  organization={IEEE}
}

@article{mohammadi2026symbolicphasor,
  title={SymbolicPhasor: Power System Phasor Estimation via Deep Symbolic Regression},
  author={Mohammadi, Sina and Su, Wencong},
  journal={arXiv preprint arXiv:2608.08552},
  year={2026}
}

@ARTICLE{5235740,
  author={Brahma, Sukumar M. and De Leon, Phillip L. and Kavasseri, Rajesh G.},
  journal={IEEE Transactions on Power Delivery}, 
  title={Investigating the Option of Removing the Antialiasing Filter From Digital Relays}, 
  year={2009},
  volume={24},
  number={4},
  pages={1864-1868},
  doi={10.1109/TPWRD.2009.2028802}}

@ARTICLE{10158378,
  author={Saleh, Saleh A. and Ozkop, Emre and Al-Durra, Ahmed and Hill, Theodore and Meng, Julian and Valdes, Marcelo E.},
  journal={IEEE Transactions on Industry Applications}, 
  title={On the Assessment of Sampling Rate Impacts on Responses of Digital Protective Relays}, 
  year={2023},
  volume={59},
  number={5},
  pages={5420-5431},
  doi={10.1109/TIA.2023.3287477}}

@ARTICLE{85841,
  author={Sachdev, M.S. and Nagpal, M.},
  journal={IEEE Transactions on Power Delivery}, 
  title={A recursive least error squares algorithm for power system relaying and measurement applications}, 
  year={1991},
  volume={6},
  number={3},
  pages={1008-1015},
  doi={10.1109/61.85841}}

@ARTICLE{5109874,
  author={Dashti, Hamed and Sanaye-Pasand, Majid and Davarpanah, Mahdi},
  journal={IEEE Transactions on Power Delivery}, 
  title={Fast and Reliable CT Saturation Detection Using a Combined Method}, 
  year={2009},
  volume={24},
  number={3},
  pages={1037-1044},
  doi={10.1109/TPWRD.2009.2022666}}

@article{phadke2009computer,
  title={Computer relaying for power systems},
  author={Phadke, Arun G and Thorp, James S},
  journal={New York, NY,},
  year={2009}
}

@article{liu2026analytical,
  title={An Analytical Phasor Estimation Method for Transient Currents Considering Double Decaying DC Offsets},
  author={Liu, Zhongping and Zou, Yizhuo and Zhao, Haoyu and Han, Jun and Liu, Yu},
  journal={IET Generation, Transmission \& Distribution},
  volume={20},
  number={1},
  pages={e70309},
  year={2026},
  publisher={Wiley Online Library}
}

@article{bjorck1970solution,
  title={Solution of Vandermonde systems of equations},
  author={Bj{\"o}rck, {\AA}ke and Pereyra, Victor},
  journal={Mathematics of Computation},
  volume={24},
  number={112},
  pages={893--903},
  year={1970}
}

\end{document}